\documentclass[12pt]{article}
\usepackage{cite}
\usepackage{amssymb,amsthm}
\usepackage[fleqn]{amsmath}
\usepackage[pdftex]{graphicx}

\newcommand{\be}{\begin{eqnarray}}
\newcommand{\ee}{\end{eqnarray}}
\newcommand{\bez}{\begin{eqnarray*}}
\newcommand{\eez}{\end{eqnarray*}}

\newcommand{\cA}{\mathcal{A}}

\newcommand{\bbC}{\mathbb{C}}

\newcommand{\bbS}{\mathbb{S}}

\renewcommand{\d}{\mathrm{d}}
\newcommand{\bd}{\bar{\mathrm{d}}}

\newcommand{\bsy}{\boldsymbol}

\theoremstyle{plain}
\newtheorem{theorem}{Theorem}[section]

\theoremstyle{definition}
\newtheorem{remark}[theorem]{Remark}
\newtheorem{example}[theorem]{Example}

\begin{document}

\title{ \textbf{Generalized Volterra lattices: \\ binary Darboux transformations and self-consistent sources} }

\author{
  \sc{F. M\"uller-Hoissen}$^a$, \sc{O. Chvartatskyi}$^{a,b}$ and \sc{K. Toda}$^c$ \\
  \small
 $^a$ Max-Planck-Institute for Dynamics and Self-Organization,
         G\"ottingen, Germany \\
  \small 
 $^b$ Mathematisches Institut, Georg-August-Universit\"at G\"ottingen, Germany \\
  \small
 $^c$ Department of Mathematical Physics, Toyama Prefectural University, Toyama, Japan
}

\date{}

\maketitle

\begin{abstract}
We study two families of matrix versions of generalized Volterra (or Bogoyavlensky) lattice equations. 
For each family, the equations arise as reductions of a partial differential-difference equation in one 
continuous and two discrete variables, which is a realization of a general integrable equation in bidifferential 
calculus. This allows to derive a binary Darboux transformation and also self-consistent source extensions 
via general results of bidifferential calculus. Exact solutions are constructed from the simplest seed solutions. 
\end{abstract}

\section{Introduction}
The famous \emph{Volterra lattice equation} \cite{Moser75,Kac+vMoer75,Mana75,Wada76,Hiro+Sats76,Kaji+Wada98} is
\be
    V_t = V \, V_{(1)} - V_{(-1)} \, V  \, ,  \label{V_lattice_eq}
\ee
here generalized to a matrix variable $V$. $V_t$ denotes the derivative of $V$ with respect to the continuous variable $t$, 
and $V_{(\pm 1)}$ are the values at the adjacent lattice sites.  
(\ref{V_lattice_eq}) is a member of the family 
\be
    ( V V_{(1)} \cdots V_{(k-1)} )_t = \left\{ \begin{array}{l} 
         V \cdots V_{(l-1)} - V_{(k-l)} \cdots V_{(k-1)}  \\
         V_{(-1)}^{-1} \cdots V_{(l)}^{-1} - V_{(k-l-1)}^{-1} \cdots V_{(k)}^{-1} 
         \end{array} \right.
         \, \mbox{if} \; \begin{array}{l} l >0 \\ l <0 \end{array}   \label{genV-lattice_eqs}
\ee
of \emph{generalized Volterra lattice equations}. Here $k>0$ and $l$ are integers and 
we recover (\ref{V_lattice_eq}) for $k=1$ and $l=2$. The family of equations (\ref{genV-lattice_eqs}) 
has been explored, mostly only in the scalar case, in 
\cite{Nari82,Bogo88a,Bogo88b,Bogo89,Bogo91,Hu+Bull97,Hu+Clar02,Hika+Inou99,DMH02LV,Wang12,Suri03,Svin11,Berk+Igon16,Adle16}, 
for example. For $k=1$ and $l=-1$,  (\ref{genV-lattice_eqs}) 
yields the \emph{modified Volterra lattice equation}
\be
     U_t = U \, (U_{(1)} - U_{(-1)}) \, U     \label{modV_lattice_eq}
\ee
in terms of $U = V^{-1}$. In Section~\ref{sec:Volterra1} we show that there is an integrable semi-discrete chiral model 
(see (\ref{sdcm})), for an invertible matrix variable $g$ depending on one continuous and two discrete 
variables, which admits the reductions
\be
     (g \, g_{(k)}^{-1})_t = g \, g_{(l)}^{-1} - (g \, g_{(l)}^{-1})_{(k-l)} \, .  \label{g_V}
\ee
Via 
\be
     V = g \, g_{(1)}^{-1} \, ,    \label{V_from_g} 
\ee
(\ref{g_V}) implies the corresponding generalized Volterra lattice equation (\ref{genV-lattice_eqs}). 

Using the framework of bidifferential calculus \cite{DMH00a,DMH08bidiff}, we construct binary Darboux transformations for the 
above mentioned semi-discrete chiral model and its reductions (\ref{g_V}). Moreover, we derive self-consistent 
source extensions (see the references cited in \cite{CDMH16}) of these equations, 
following the general construction developed in \cite{CDMH16}.\footnote{A  
self-consistent source extension of the scalar modified Volterra lattice equation appeared in \cite{Yu09}.} 
We also refer to the latter publication for a representative list of references concerning  
this way of extending an integrable equation to an equation with additional ``source terms'', which 
involve new dependent variables governed by equations that also involve the variable of the original equation.

A second family of equations generalizing the Volterra lattice equation is given by 
\be
  \Big( \sum_{i=0}^{k-1} W_{(i)} \Big)_t = \sum_{i=0}^{l-1} W_{(i)}  \sum_{i=0}^{k-1} W_{(i)} 
       - \sum_{i=0}^{k-1} W_{(i)} \sum_{i=k-l}^{k-1} W_{(i)}  \, . \label{genW_eqs}
\ee
Such equations have been explored in \cite{Itoh87,Bogo88a,Bogo88b,Bogo89,Bogo91,Suri99,DMH02LV,Suri03,QTLWLL11}, 
mostly in the scalar case. For $k=1$ this reduces to
\bez
  W_t = \sum_{i=1}^{l-1} W_{(i)} \, W - W \, \sum_{i=1}^{l-1} W_{(-i)} \, ,
\eez
which for $l=2$ is the Volterra lattice equation in the form $W_t = W_{(1)} W - W W_{(-1)}$. 
There is a differential-difference equation (see (\ref{2+1_varphi_eq})), for a variable $\varphi$ depending 
on one continuous and two discrete variables, which admits reductions to
\be
     (\varphi_{(k)} - \varphi)_t = (\varphi_{(l)} - \varphi)(\varphi_{(k)} - \varphi +I)
   -   (\varphi_{(k)} - \varphi + I) ( \varphi_{(l)} - \varphi)_{(k-l)} \, ,   \label{varphi_preW_eqs}
\ee
where $I$ is the identity matrix, and we observe that, in terms of 
\be
     W = \varphi_{,1} - \varphi + \frac{1}{k} \, I \, ,   \label{W_from_varphi} 
\ee
the latter becomes (\ref{genW_eqs}). Using results of \cite{CDMH16}, in the framework of
bi\-differential calculus, we construct binary Darboux transformations and self-consistent source extensions 
for these equations. 

The equations (\ref{genV-lattice_eqs}) and (\ref{genW_eqs}), for $k=1$, are also known as \emph{Bogoyavlensky lattices} 
(cf. \cite{Suri03}).  

In Section~\ref{sec:Volterra1} we present a bidifferential calculus formulation for the semi-discrete 
chiral model. Section~\ref{subsec:scs_in_bdc} recalls from \cite{CDMH16} some general results concerning 
binary Darboux transformations and self-consistent source extensions of integrable equations in this framework. 
In Section~\ref{subsec:sdcm} we derive self-consistent source extensions of the semi-discrete 
chiral model equation and construct corresponding exact solutions. 
In Section~\ref{subsec:scsV1} we derive self-consistent source extensions of the first family of 
generalized Volterra lattice equations as reductions of the self-consistent source extensions of 
the semi-discrete chiral model. We also present infinite families of exact solutions of these equations. 
Section~\ref{sec:Volterra2} contains a corresponding treatment of the second family of 
generalized Volterra lattice equations. 
Section~\ref{sec:conclusions} contains some concluding remarks.

\section{An integrable semi-discrete chiral model and the first family of generalized Volterra lattices}
\label{sec:Volterra1}
Let $\boldsymbol{\Omega} = \bigoplus_{r \geq 0} \boldsymbol{\Omega}^r$ be an associative unital graded algebra over $\bbC$. 
In particular, $\cA := \boldsymbol{\Omega}^0$ is an associative unital algebra over $\bbC$ 
and $\boldsymbol{\Omega}^r$, $r \geq 1$, are $\cA$-bimodules such that 
$\boldsymbol{\Omega}^r \, \boldsymbol{\Omega}^s \subseteq \boldsymbol{\Omega}^{r+s}$.
A \emph{bidifferential calculus} is an associative unital graded algebra $\boldsymbol{\Omega}$, supplied
with two $\bbC$-linear, graded derivations $\d, \bar{\d} : \boldsymbol{\Omega} \rightarrow \boldsymbol{\Omega}$
of degree one (hence $\d \boldsymbol{\Omega}^r \subseteq \boldsymbol{\Omega}^{r+1}$,
$\bar{\d} \boldsymbol{\Omega}^r \subseteq \boldsymbol{\Omega}^{r+1}$), and such that 
$\d^2 = \bd^2 = \d \bd + \bd \d = 0$.   
In this work we choose, as in \cite{CDMH16}, the graded algebra $\boldsymbol{\Omega}$ to be of the form
$\boldsymbol{\Omega} = \cA \otimes \bsy{\Lambda}$ with the exterior (Grassmann) algebra 
$\bsy{\Lambda} = \bigoplus_{i=0}^2 \bsy{\Lambda}^i$ of the vector space $\bbC^2$.    
It is then sufficient to define $\d$ and $\bd$ on $\cA$, since they extend to $\boldsymbol{\Omega}$ 
in a straightforward way, treating the elements of $\bsy{\Lambda}$ as ``constants''. Moreover, 
$\d$ and $\bd$ extend to matrices over $\boldsymbol{\Omega}$. 
We choose a basis $\xi_1,\xi_2$ of $\bsy{\Lambda}^1$.

Next we specify the bidifferential calculus as follows. 
Let $\cA_0$ be the space of complex functions of one continuous and two discrete variables, and 
$\bbS_1,\bbS_2$ corresponding shift operators. 
We extend $\cA_0$ to $\cA = \cA_0[\bbS_1^{\pm 1},\bbS_2^{\pm 1}]$ and define $\d$ 
and $\bar{\d}$ on $\cA$ via 
\be
    \d f = - [\bbS_1 \bbS_2^{-1} , f] \, \xi_1 + f_t \, \xi_2 \, , \qquad
    \bd f = [\bbS_1 , f] \, \xi_1 + [\bbS_2 , f] \, \xi_2 \, .    \label{sdcm_bidiff}
\ee
In the following we will use the notation
\bez
    f_{,i} = \bbS_i \, f \, \bbS_i^{-1} \, , \quad
    f_{,-i} = \bbS_i^{-1} \, f \, \bbS_i \, , \qquad i=1,2 \, .
\eez 
 From the equation
\be
     \d [ (\bd g) \, g^{-1} ] = 0 \, ,  \label{g_eq}
\ee
for an invertible $g \in \mathrm{Mat}(m,m,\cA)$, i.e., an $m \times m$ matrix with entries in $\cA$, 
quite a number of integrable equations can be derived and it comes along with 
universal solution-generating methods \cite{DMH08bidiff,DMH13SIGMA}. 
With the above choice of bidifferential calculus, it takes the form
\be
     (g \, g_{,1}^{-1})_t + (g \, g_{,2}^{-1})_{,1,-2} - g \, g_{,2}^{-1} = 0 \, ,  \label{sdcm}
\ee
and we can restrict $g$ to $\mathrm{Mat}(m,m,\cA_0)$. This is an integrable semi-discrete chiral model 
equation in one continuous and two discrete variables. The fact that (\ref{sdcm}) is a realization  
of (\ref{g_eq}) is an expression of its integrability. A solution generating method for (\ref{g_eq}), 
recalled in Section~\ref{subsec:scs_in_bdc}, then specializes to (\ref{sdcm}).
In the scalar case ($m=1$), writing $g= e^{-u}$, (\ref{sdcm}) becomes
\bez
    (u_{,1} - u)_t = e^{u_{,2} - u_{,1}} + e^{(u_{,2}-u_{,1})_{,-2}}   \, .
\eez

Specializing the above shift operators in terms of a shift operator $\bbS$ as follows,
\be
     \bbS_1 = \bbS^k \, , \qquad \bbS_2 = \bbS^l \, ,    \label{V_shifts}
\ee
with fixed integers $k>0$ and $l$, and introducing  
\bez
      g_{(k)} = \bbS^k g \, \bbS^{-k} \, ,
\eez 
then (\ref{sdcm}) becomes (\ref{g_V}), which in turn implies the generalized Volterra lattice equation 
(\ref{genV-lattice_eqs}).

\begin{remark}
\label{rem:d,bd_exchange_for_g-eq}
By exchanging $\d$ and $\bd$ in (\ref{g_eq}), we obtain instead of (\ref{sdcm}) the following equation,
\bez
    (g_t \, g^{-1})_{,1} - g_t \, g^{-1} = g_{,2} \, g_{,1}^{-1} - (g_{,2} \, g_{,1}^{-1})_{,-2} \, .
\eez
This is equivalent to (\ref{sdcm}) via $g \mapsto g^{-1}$. 
\end{remark}

\subsection{Binary Darboux transformations and self-consistent source extensions for (\ref{g_eq})}
\label{subsec:scs_in_bdc}
Let $\Delta,\Gamma \in \mathrm{Mat}(n,n,\cA)$ 
and $\alpha,\beta$ be $n \times n$ matrices of elements of $\boldsymbol{\Omega}^1$, subject to
\be
  &&  \bd \Delta + [\alpha , \Delta] = (\d \Delta) \, \Delta \, , \qquad  
      \bd \alpha + \alpha^2 = (\d \alpha) \, \Delta \, , \nonumber \\ 
  &&  \bd \Gamma - [\beta , \Gamma] = \Gamma \, \d \Gamma  \, , \hspace{1.5cm}
      \bd \beta - \beta^2 = \Gamma \, \d \beta \, .   \label{Delta,alpha,Gamma,beta_eqs}
\ee
We introduce
\be
    \gamma = \bd \omega - (\d \omega) \, \Delta + (\d \Gamma) \, \omega - \beta \, \omega 
              - \omega \, \alpha \, ,
                                   \label{gamma_in_terms_of_omega}
\ee
with $\omega \in \mathrm{Mat}(n,n,\cA)$, and impose the constraint
\be
      \Gamma \, \omega = \omega \, \Delta \, .  \label{c=0_constraint}
\ee
By use of (\ref{Delta,alpha,Gamma,beta_eqs}), the latter implies
\bez
      \Gamma \, \gamma = \gamma \, \Delta \, . 
\eez
We recall from \cite{CDMH16} the following result. 

\begin{theorem}
\label{thm:main_g}
Let $g_0 \in \mathrm{Mat}(m,m,\cA)$ be an invertible solution of (\ref{g_eq}) and $\Delta, \Gamma, \alpha, \beta$ 
solutions of (\ref{Delta,alpha,Gamma,beta_eqs}). 
Let $\theta \in \mathrm{Mat}(m,n,\cA)$, $\eta \in \mathrm{Mat}(n,m,\cA)$ solve the linear equations 
\be
  \bd \theta = (\bd g_0) \, g_0^{-1} \, \theta + (\d \theta) \, \Delta + \theta \, \alpha \, ,   \quad
  \bd \eta = - \eta \, (\bd g_0) \, g_0^{-1} + \Gamma \, \d \eta + \beta \, \eta \, . \quad  \label{theta,eta_eqs}
\ee
Furthermore, let $\Omega$ be a solution of the linear equations 
\be
  \Gamma \, \Omega - \Omega \, \Delta = \eta \, \theta \, , \quad
  \bd \Omega = (\d \Omega) \, \Delta - (\d \Gamma) \, \Omega 
     + \beta \, \Omega + \Omega \, \alpha + (\d \eta) \, \theta + \gamma \, ,  \quad \label{scs_Omega_eqs}
\ee
where $\gamma$ is given by (\ref{gamma_in_terms_of_omega}) in terms of some 
$\omega \in \mathrm{Mat}(n,n,\cA)$ satisfying (\ref{c=0_constraint}). 
If $\Gamma$ and $\Omega$ are invertible, then
\be
    g = (I - \theta \, \Omega^{-1} \, \Gamma^{-1} \eta ) \, g_0 \, , \qquad
    q = \theta \, \Omega^{-1} \, , \qquad 
    r = \Omega^{-1} \, \eta  \, ,                \label{g,q,r}
\ee
where $I$ is the $m \times m$ identity matrix, solves 
\be
       \d [ (\bd g) \, g^{-1} ] = \d (q \, \gamma \, \Delta^{-1} \, r )   \label{scs_g_eq} 
\ee
and 
\be
   && \bd q = (\bd g) \, g^{-1} \, q + \d( q \, \Gamma) - q \, \beta 
              - q \, \gamma \, ( \Omega^{-1} + \Delta^{-1} r q ) \, ,  \label{scs_q_eq} \\
   && \bd r = - r \, (\bd g) \, g^{-1} + \d( \Delta \, r )  
        - \alpha \, r  - ( \Omega^{-1} - r q \, \Gamma^{-1} ) \, \gamma \, r  \, . \label{scs_r_eq}
\ee
\hfill $\Box$
\end{theorem}

If $\gamma =0$, the theorem expresses a binary Darboux transformation for 
the equation (\ref{g_eq}). A non-vanishing $\gamma$ switches ``sources'' on, since  
(\ref{scs_g_eq}) has the form of a source-extended version of (\ref{g_eq}). More precisely,  
integrable equations with self-consistent sources that appeared in the literature (see the references in \cite{CDMH16}) 
are recovered by reducing (\ref{scs_q_eq}) and (\ref{scs_r_eq}) to equations that do not involve $\Omega$. This is achieved 
by restrictions on $\omega$ and by disregarding half of the set of equations obtained from 
(\ref{scs_q_eq}) and (\ref{scs_r_eq}), cf. \cite{CDMH16}. In Section~\ref{subsec:sdcm} we will exploit 
this for the semi-discrete chiral model (\ref{sdcm}).

As formulated above, the number of source terms is $n$, since $q$ and $r$ have both $n$ components. But $q$ and $r$ enter 
the right hand side of (\ref{scs_g_eq}) sandwiching an expression linear in $\omega$. Hence, if the latter matrix has rank 
$N < n$, then only $N$ source terms are present. In this way, $n$-soliton solutions of a system with $N$ sources 
can be constructed.

\subsection{Binary Darboux transformations and self-consistent source extensions for the semi-discrete chiral model equation}
\label{subsec:sdcm}
Using the bidifferential calculus determined by (\ref{sdcm_bidiff}), by inspection of the equations 
in Section~\ref{subsec:scs_in_bdc} we are guided to write
\be
  && q = \tilde{q} \, \bbS_2 \, , \quad
     r = \bbS_2 \, \tilde{r} \, , \quad
     \Delta = \Gamma = \bbS_2 \, , \nonumber \\
  &&  \alpha = -(P+2 I) \, \bbS_1 \, \xi_1 - \bbS_2 \, \xi_2 \, , \quad
     \beta = (Q + 2 I) \, \bbS_1 \, \xi_1 + \bbS_2 \, \xi_2  \, , \quad  \nonumber \\
  && \Omega = \tilde{\Omega} \, \bbS_2^{-1} \, , \quad 
     \omega = \tilde{\omega} \, \bbS_2^{-1} \, , \quad 
     \gamma = \gamma_1 \, \bbS_1 \, \bbS_2^{-1} \, \xi_1 + \gamma_2 \, \xi_2   \, .  \label{sdcm_redefinitions}
\ee
Then it turns out that $\tilde{q}, \tilde{r}, P,Q, \gamma_1, \gamma_2, \tilde{\omega}$ 
and $\tilde{\Omega}$ can be restricted to be matrices over $\cA_0$ instead of $\cA$ (with the algebras chosen  
in the beginning of this section). The equations (\ref{Delta,alpha,Gamma,beta_eqs}) then take the form
\be
    P_{,2} = P \, , \qquad P_t = 0 \, , \qquad
    Q_{,2} = Q \, , \qquad Q_t = 0 \, .            \label{P,Q_etc_eqs}
\ee
The constraint (\ref{c=0_constraint}) becomes   
\be
     \tilde{\omega}_{,2} = \tilde{\omega} \, ,   \label{sdcm_c=0}
\ee
and (\ref{gamma_in_terms_of_omega}) leads to
\be
  \gamma_1 = \tilde{\omega} \, P - Q \, \tilde{\omega}_{,1}   \, , \qquad
  \gamma_2 = - \tilde{\omega}_t \, .   \label{sdcm_gamma}
\ee
(\ref{scs_g_eq}) takes the form
\be
    \hspace*{-.4cm}
    (g \, g_{,1}^{-1})_t + (g \, g_{,2}^{-1})_{,1,-2} - g \, g_{,2}^{-1}
  = (\tilde{q}_{,-2} \, \gamma_2 \, \tilde{r})_{,2}
    - (\tilde{q}_{,-2} \, \gamma_2 \, \tilde{r})_{,1} - ( \tilde{q} \, \gamma_1 \, \tilde{r}_{,1} )_t \, .  \label{V_scs_g}
\ee
(\ref{scs_q_eq}) splits into the two equations
\be
  \tilde{q}_{,1,-2} 
     &=& - \tilde{q} \, Q - g \, g^{-1}_{,1} \, \tilde{q}_{,1} 
         - \tilde{q} \, \gamma_1 \, ( \tilde{\Omega}_{,2}^{-1} + \tilde{r} \tilde{q} )_{,1} \, , \nonumber \\
  \tilde{q}_t &=& g \, g^{-1}_{,2} \tilde{q}_{,2} 
     + \tilde{q} \, \gamma_2 \, ( \tilde{\Omega}^{-1}_{,2} + \tilde{r} \tilde{q} )_{,2} \, .  \label{V_scs_q}
\ee
Correspondingly, from (\ref{scs_r_eq}) we obtain
\be
  \tilde{r}_{,2} 
    &=& - P \, \tilde{r}_{,1}
      - \tilde{r} \, g \, g^{-1}_{,1} + ( \tilde{\Omega}^{-1} - \tilde{r} \tilde{q} ) \, \gamma_1 \, \tilde{r}_{,1}  
              \, ,  \nonumber \\
  \tilde{r}_t &=& - (\tilde{r} \, g)_{,-2} \, g^{-1} 
         + (\tilde{\Omega}^{-1} - \tilde{r} \tilde{q} )_{,-2} \, \gamma_2 \, \tilde{r} \, .  \label{V_scs_r}
\ee

According to \cite{CDMH16}, we obtain self-consistent source extensions of the semi-discrete chiral model (\ref{sdcm}) 
by reducing (\ref{V_scs_q}) and (\ref{V_scs_r}) to equations that do not contain $\tilde{\Omega}$. 
This is achieved by setting either $\gamma_1$ or $\gamma_2$ to zero and disregarding one from each pair of 
equations (\ref{V_scs_q}) and (\ref{V_scs_r}).

\begin{enumerate}
\item $\gamma_1 =0$, i.e., $Q \, \tilde{\omega}_{,1} = \tilde{\omega} \, P$. 
\bez 
  &&   (g \, g_{,1}^{-1})_t + (g \, g_{,2}^{-1})_{,1,-2} - g \, g_{,2}^{-1}
  =  (\tilde{q}_{,-2} \, \gamma_2 \, \tilde{r})_{,2}
    - (\tilde{q}_{,-2} \, \gamma_2 \, \tilde{r})_{,1} \, , \\
  && \tilde{q}_{,1,-2} = - \tilde{q} \, Q - g \, g^{-1}_{,1} \, \tilde{q}_{,1}  \, ,  \qquad
     \tilde{r}_{,2} = - P \, \tilde{r}_{,1} - \tilde{r} \, g \, g^{-1}_{,1}   \, .     
\eez
As a consequence of $Q \, \tilde{\omega}_{,1} = \tilde{\omega} \, P$, (\ref{P,Q_etc_eqs}) and the expression for 
$\gamma_2$ in (\ref{sdcm_gamma}), we have $Q \, \gamma_{2,1} = \gamma_2 \, P$. This allows to absorb $\gamma_2$ 
by introducing
\be
     \hat{q} = \tilde{q} \, \gamma_2 \, .   \label{hatq_scs_V1}
\ee
The above system then reads
\be 
  &&   (g \, g_{,1}^{-1})_t + (g \, g_{,2}^{-1})_{,1,-2} - g \, g_{,2}^{-1}
  =  (\hat{q}_{,-2} \, \tilde{r})_{,2}
    - (\hat{q}_{,-2} \, \tilde{r})_{,1} \, , \nonumber \\
  && \hat{q}_{,1,-2} = - \hat{q} \, P - g \, g^{-1}_{,1} \, \hat{q}_{,1} \, , \qquad 
     \tilde{r}_{,2} = - P \, \tilde{r}_{,1} - \tilde{r} \, g \, g^{-1}_{,1} \, .  \hspace{1cm}  \label{sdcm_scs1} 
\ee
\item $\gamma_2 =0$, i.e., $\tilde{\omega}_t =0$.
\bez 
  &&   (g \, g_{,1}^{-1})_t + (g \, g_{,2}^{-1})_{,1,-2} - g \, g_{,2}^{-1}
  = - ( \tilde{q} \, \gamma_1 \, \tilde{r}_{,1} )_t  \, , \\
  && \tilde{q}_t = g \, g^{-1}_{,2} \tilde{q}_{,2} \, , \qquad
     \tilde{r}_t = - (\tilde{r} \, g)_{,-2} \, g^{-1} \, .       
\eez
In terms of
\be
    \hat{q} = \tilde{q} \, \gamma_1 \, ,   \label{hatq_scs_V2}
\ee
using $\gamma_{1 t} = 0$ the last system takes the form
\be 
  &&   (g \, g_{,1}^{-1})_t + (g \, g_{,2}^{-1})_{,1,-2} - g \, g_{,2}^{-1}
  = - ( \hat{q} \, \tilde{r}_{,1} )_t  \, , \nonumber \\
  && \hat{q}_t = g \, g^{-1}_{,2} \hat{q}_{,2} \, , \qquad 
     \tilde{r}_t = - (\tilde{r} \, g)_{,-2} \, g^{-1} \, .    \label{sdcm_scs2}   
\ee
\end{enumerate}
The systems (\ref{sdcm_scs1}) and (\ref{sdcm_scs2}) are self-consistent source extensions of the semi-discrete 
chiral model (\ref{sdcm}). They do not involve $\tilde{\omega}$, which is, however, still relevant for 
the accompanying solution-generating method. 

\begin{remark}
Since in the first case the $t$-dependence of $\tilde{\omega}$ can be arbitrary, the system (\ref{sdcm_scs1}) 
admits solutions depending on arbitrary functions of $t$. 
\end{remark}

\subsubsection{Binary Darboux transformation}
\label{subsec:BDT}
In this subsection we elaborate Theorem~\ref{thm:main_g} for the self-consistent source extensions of the semi-discrete 
chiral model (\ref{sdcm}). 
The linear equations (\ref{theta,eta_eqs}), evaluated with the bidifferential calculus (\ref{sdcm_bidiff}) and 
using (\ref{sdcm_redefinitions}), take the form
\bez
  &&  \theta_{,-2} = - (\theta \, P)_{,-1} - g_{0,-1} \, g_0^{-1} \, \theta \, , \qquad 
     \theta_t  = g_0 \, g_{0,2}^{-1} \, \theta_{,2} \, , \\
  && \eta_{,2} = - Q \, \eta_{,1} - \eta \, g_0 \, g_{0,1}^{-1} \, , \hspace{2.05cm}
     \eta_t = - (\eta \, g_0)_{,-2} \, g_0^{-1} \, ,
\eez
and the linear equations (\ref{scs_Omega_eqs}) determining $\tilde{\Omega}$ read
\be
  && \tilde{\Omega}_{,2} - \tilde{\Omega} = \eta \, \theta \, , \nonumber \\
  && \tilde{\Omega}_{,1,-2} - (I + Q) \, \tilde{\Omega}_{,1} + \tilde{\Omega} \, P  
     + ( \eta_{,1,-2} - \eta ) \, \theta_{,1,-2}  - \gamma_1 = 0 \, , \nonumber \\
  && \tilde{\Omega}_t =  - \eta_t \, \theta - \gamma_2 \, .    \label{sdcm_Omega}
\ee
 For a given solution $g_0$ of (\ref{sdcm}), we have to find a solution of the system of linear equations for $\theta$ and $\eta$, 
and then determine a corresponding invertible solution of the linear equations for $\tilde{\Omega}$. 
Then
\be
     g = (I - \theta \, \tilde{\Omega}_{,2}^{-1} \, \eta ) \, g_0 \, , \qquad
     \tilde{q} = \theta \, \tilde{\Omega}_{,2}^{-1}  \, , \qquad
     \tilde{r} = \tilde{\Omega}^{-1} \eta    \label{sdcm_new_solution}
\ee
constitutes a solution of (\ref{V_scs_g}) - (\ref{V_scs_r}) if $P,Q$ satisfy (\ref{P,Q_etc_eqs}), 
and if $\tilde{\omega}_{,2} = \tilde{\omega}$. Via (\ref{hatq_scs_V1}), respectively (\ref{hatq_scs_V2}), 
we obtain a solution of the self-consistent source extensions (\ref{sdcm_scs1}) and (\ref{sdcm_scs2}), provided that $\tilde{\omega}$ 
obeys the respective constraint. If $\tilde{\omega}$ satisfies $\gamma_1 = \gamma_2 =0$ (cf. (\ref{sdcm_gamma})), 
then this is a binary Darboux transformation for the semi-discrete chiral model equation (\ref{sdcm}). 

\begin{example}
\label{ex:sdcm_solutions}
For constant seed $g_0$, the linear system reads
\be
  \theta_{,-2} = - (\theta \, P)_{,-1} - \theta \, , \quad 
  \theta_t = \theta_{,2} \, , \quad
  \eta_{,2} = - Q \, \eta_{,1} - \eta  \, , \quad
  \eta_t = - \eta_{,-2}  \, .   \label{sdcm_const_seed_linsys}
\ee
Choosing $P$ and $Q$ to be constant, solutions are given by 
\be
   && \theta = \sum_{a=1}^N A_a \, \Lambda_a^{-{j_1}} (-1)^{j_2} (I+\Lambda_a P)^{-j_2} \, e^{ -(I+\Lambda_a P)^{-1} \, t} \, ,  
          \nonumber \\
   && \eta = \sum_{a=1}^N e^{ (I+Q \tilde{\Lambda}_a)^{-1} \, t} \, \tilde{\Lambda}_a^{j_1} 
                   \, (-1)^{j_2} (I+Q \tilde{\Lambda}_a)^{j_2} \, B_a \, ,   \label{sdcm_cseed_theta,eta_solutions}
\ee
with constant matrices $A_a,B_a,\Lambda_a,\tilde{\Lambda}_a$, if $\Lambda_a,P$ and $\tilde{\Lambda}_a,Q$ commute, $a=1,\ldots,N$. 
Here $j_1$ and $j_2$ are independent discrete variables on which the shift operators $\bbS_1$, respectively $\bbS_2$, act. 
The corresponding solution of (\ref{sdcm_Omega}) is
\be
    \tilde{\Omega} = \sum_{a,b=1}^N e^{ (I+Q \tilde{\Lambda}_a)^{-1} \, t} \, \tilde{\Lambda}_a^{j_1} \, (I+Q \tilde{\Lambda}_a)^{j_2}
    X_{ab} \, \Lambda_b^{-{j_1}} (I+\Lambda_b P)^{-j_2} \, e^{ -(I+\Lambda_b P)^{-1} \, t} + \tilde{\omega} \, , \hspace*{-1cm}\nonumber \\
                       \label{sdcm_cseed_Omega_solution}
\ee
where, for $a,b=1,\ldots,N$, $X_{ab}$ is a constant $n \times n$ matrix solution of the \emph{Stein equation}
\bez
    (I+Q \tilde{\Lambda}_a) \, X_{ab} \, (I+\Lambda_b P)^{-1} - X_{ab} = B_a \, A_b \, .
\eez
Now (\ref{sdcm_new_solution}), with constant $g_0$, and (\ref{hatq_scs_V1}), respectively (\ref{hatq_scs_V2}), provides us with a 
set of solutions of the above self-consistent source 
extensions of the semi-discrete chiral model, if $\tilde{\omega}$ satisfies the respective condition. 
If $\tilde{\omega}$ is constant and satisfies $\tilde{\omega} P = Q \, \tilde{\omega}$, 
then $\gamma_1 = \gamma_2 =0$ and we have solutions of the semi-discrete chiral model equation (\ref{sdcm}). 
The above solutions have been obtained by assuming that the two shift operators are independent. Hence they cannot be used 
conveniently to obtain solutions of Volterra lattices, where (\ref{V_shifts}) holds. 
\end{example}

\subsection{Self-consistent source extensions for the first family of generalized Volterra lattice equations and exact solutions}
\label{subsec:scsV1}
Using the reduction (\ref{V_shifts}), the systems (\ref{sdcm_scs1}) and (\ref{sdcm_scs2}) 
yield the following self-consistent source extensions of (\ref{g_V}). In this subsection we will assume 
that $P$ and $Q$ are constant and that $\tilde{\omega}$ only depends on $t$.\footnote{More generally, 
these quantities are still allowed to depend on the discrete variable in a periodic way: 
$P_{(l)} = P$, $Q_{(l)} = Q$, $\tilde{\omega}_{(l)} = \tilde{\omega}$. }

\begin{enumerate}
\item $Q \, \tilde{\omega} = \tilde{\omega} \, P$. 
Then (\ref{sdcm_scs1}) becomes
\be 
   &\hspace{-1cm}&   (g \, g_{(k)}^{-1})_t + (g \, g_{(l)}^{-1})_{(k-l)} - g \, g_{(l)}^{-1}
  =  (\hat{q}_{(-l)} \, \tilde{r})_{(l)}
    - (\hat{q}_{(-l)} \, \tilde{r})_{(k)} \, , \nonumber \\
  &\hspace{-1cm}& \hat{q}_{(k-l)} + \hat{q} \, P + g \, g^{-1}_{(k)} \, \hat{q}_{(k)}  = 0 \, , \quad
     \tilde{r}_{(l-k)} + P \, \tilde{r} + (\tilde{r} \, g)_{(-k)} \, g^{-1} = 0  \, .  \hspace{1cm}  \label{gVscs1} 
\ee
Using (\ref{V_from_g}) this implies
\be
  && \hspace*{-.8cm} ( V V_{(1)} \cdots V_{(k-1)} )_t 
     - \left\{ \begin{array}{l} 
         V \cdots V_{(l-1)} - V_{(k-l)} \cdots V_{(k-1)}  \\
         V_{(-1)}^{-1} \cdots V_{(l)}^{-1} - V_{(k-l-1)}^{-1} \cdots V_{(k)}^{-1} 
         \end{array} \right.
         \mbox{if} \, \begin{array}{l} l >0 \\ l <0 \end{array}   \nonumber \\
  && \hspace{2.4cm} = ( \hat{q} \, \tilde{r}_{(l)} )_{(k-l)} - \hat{q} \, \tilde{r}_{(l)}  \, , \nonumber \\
  && \hspace*{-.8cm}
      \hat{q}_{(k-l)} + \hat{q} \, P = - V V_{(1)} \cdots V_{(k-1)} \, \hat{q}_{(k)} \, , \nonumber \\
  && \hspace*{-.8cm}   
     \tilde{r}_{(l)} + P \, \tilde{r}_{(k)} = - \tilde{r} \, V V_{(1)} \cdots V_{(k-1)}  \, .  \label{scs_V1}
\ee
\item 
$\tilde{\omega}_t = 0$. 
Then (\ref{sdcm_scs2}) yields
\be 
  &&   (g \, g_{(k)}^{-1})_t + (g \, g_{(l)}^{-1})_{(k-l)} - g \, g_{(l)}^{-1}
  = - ( \hat{q} \, \tilde{r}_{(k)} )_t  \, , \nonumber \\
  && \hat{q}_t = g \, g^{-1}_{(l)} \hat{q}_{(l)} \, , \qquad 
     \tilde{r}_t = - (\tilde{r} \, g)_{(-l)} \, g^{-1} \, .    \label{gVscs2}   
\ee
Using (\ref{V_from_g}) this leads to
\be
  && ( V V_{(1)} \cdots V_{(k-1)} )_t 
      - \left\{ \begin{array}{l} 
         V \cdots V_{(l-1)} - V_{(k-l)} \cdots V_{(k-1)}  \\
         V_{(-1)}^{-1} \cdots V_{(l)}^{-1} - V_{(k-l-1)}^{-1} \cdots V_{(k)}^{-1} 
         \end{array} \right.
         \mbox{if}  \begin{array}{l} l >0 \\ l <0 \end{array}   \nonumber \\
  && \hspace{2.4cm} = - ( \hat{q} \, \tilde{r}_{(k)} )_t  \, , \qquad  \nonumber \\ 
  && \hat{q}_t = \left\{ \begin{array}{l} 
         V \cdots V_{(l-1)} \, \hat{q}_{(l)}  \\
         V_{(-1)}^{-1} \cdots V_{(l)}^{-1} \, \hat{q}_{(l)}  \end{array} \right.
         \mbox{if} \begin{array}{l} l >0 \\ l <0 \end{array}  \, , \nonumber \\   
  && \tilde{r}_t = - \tilde{r}_{(-l)} \, \left\{ \begin{array}{l} 
         V_{(-l)} \cdots V_{(-1)} \\
         V_{(-l-1)}^{-1} \cdots V^{-1}  \end{array} \right.
         \mbox{if}  \begin{array}{l} l > 0 \\ l < 0 \end{array}  \, .  \hspace{1cm}     \label{scs_V2}  
\ee  
 For $k=1$ and $l<0$, in terms of $U = V^{-1}$ this reads
\be
  && U_t - U \, [ U_{(-l)} \cdots U_{(1)} - U_{(-1)} \cdots U_{(l)} ] \, U = U \, ( \hat{q} \, \tilde{r}_{(1)} )_t \, U \, , 
       \hspace{1cm}  \nonumber \\ 
  && \hat{q}_t = U_{(-1)} \cdots U_{(l)} \, \hat{q}_{(l)}
     \, , \qquad
     \tilde{r}_t = - \tilde{r}_{(-l)} \, U_{(-l-1)} \cdots U \, .  \label{gmVolterra_eqs} 
\ee
\end{enumerate}

\subsubsection{Exact solutions with constant seed}
\label{subsec:V1_exact_sol}
 For constant seed $g_0$, we obtain from (\ref{sdcm_const_seed_linsys}) via (\ref{V_shifts}) 
the linear system\footnote{It should be noticed that in the derivation of (\ref{sdcm_const_seed_linsys}) 
and other equations in Section~\ref{subsec:sdcm}, we did \emph{not} assume that the shift operators $\bbS_1$ 
and $\bbS_2$ are independent. Only in Example~\ref{ex:sdcm_solutions} this is assumed. } 
\bez
   \theta_t = \theta_{(l)} = - \theta_{(l-k)} P - \theta  \, , \qquad 
   \eta_t = - \eta_{(-l)} = Q \, \eta_{(k-l)} + \eta  \, , 
\eez
where $P$ and $Q$ are constant. Writing
\be
     P = - ( \Lambda^{k} + \Lambda^{k-l} ) \, , \qquad 
     Q = - ( \tilde{\Lambda}^k + \tilde{\Lambda}^{k-l} ) \, ,    \label{V_P,Q-Lambda}
\ee
with constant $n \times n$ matrices $\Lambda$ and $\tilde{\Lambda}$, solutions are given by
\bez
    \theta = \sum_{\Lambda \in \mathfrak{R}(P)} A_\Lambda \, e^{ \Lambda^l \, t} \, \Lambda^j 
                   \, , \qquad
    \eta = \sum_{\tilde{\Lambda} \in \mathfrak{R}(Q)} e^{-\tilde{\Lambda}^l \, t} \, 
              \tilde{\Lambda}^{-j} \, B_{\tilde{\Lambda}}   \, ,
\eez
where $\mathfrak{R}(P)$, respectively $\mathfrak{R}(Q)$, is a set of distinct roots of (\ref{V_P,Q-Lambda}), regarded 
as a polynomial matrix equation for $\Lambda$, respectively $\tilde{\Lambda}$. Here $j$ is the discrete variable on which 
the shift operator $\bbS$ acts. $A_\Lambda$, $\Lambda \in \mathfrak{R}(P)$,
and $B_{\tilde{\Lambda}}$, $\tilde{\Lambda} \in \mathfrak{R}(Q)$, are constant matrices. 

The equations for $\tilde{\Omega}$ are then solved by
\bez
    \tilde{\Omega} = \sum_{\Lambda \in \mathfrak{R}(P),\tilde{\Lambda} \in \mathfrak{R}(Q)}
    e^{-\tilde{\Lambda}^l \, t} \, \tilde{\Lambda}^{-j} \, X_{\tilde{\Lambda},\Lambda} \, 
       e^{ \Lambda^l \, t} \, \Lambda^j 
            + \tilde{\omega} \, ,
\eez
where the constant matrices $X_{\tilde{\Lambda},\Lambda}$ have to solve the Stein equations
\bez
  \tilde{\Lambda}^{-l} \, X_{\tilde{\Lambda},\Lambda} \, \Lambda^l - X_{\tilde{\Lambda},\Lambda} 
       = B_{\tilde{\Lambda}} \, A_{\Lambda} \qquad \quad 
         \forall \, \Lambda \in \mathfrak{R}(P), \, \tilde{\Lambda} \in \mathfrak{R}(Q) \, .     
\eez
Then
\bez
     g = (I - \theta \, \tilde{\Omega}_{(l)}^{-1} \, \eta) \, g_0  \, , \qquad
     \hat{q} = \theta \, \tilde{\Omega}_{(l)}^{-1} \gamma_i  \, , \qquad
     \tilde{r} = \tilde{\Omega}^{-1} \eta 
\eez
solves (\ref{gVscs1}) if $i=2$ and (\ref{gVscs2}) if $i=1$. The corresponding constraint for $\tilde{\omega}$ 
has to be satisfied, of course. Via (\ref{V_from_g}), this yields solutions of the Volterra lattice equation 
(\ref{scs_V1}), respectively (\ref{scs_V2}), with self-consistent sources. 

The only hurdle is (\ref{V_P,Q-Lambda}), which should be read as follows. Choose any constant $n \times n$ 
matrix $\Lambda$ and define $P$ by (\ref{V_P,Q-Lambda}). Then at least one more solution of (\ref{V_P,Q-Lambda})
with the same $P$ has to be found. Correspondingly for the second equation in (\ref{V_P,Q-Lambda}). 
See the examples treated in the following two subsections. 

\begin{remark}
If $\Lambda = \mathrm{diag}(\lambda_1,\ldots,\lambda_n)$ and 
$\tilde{\Lambda} = \mathrm{diag}(\tilde{\lambda}_1,\ldots,\tilde{\lambda}_n)$, 
the solution matrix $X_{\tilde{\Lambda},\Lambda}$ of the above Stein equation has entries
\bez
      (X_{\tilde{\Lambda},\Lambda})_{ij} 
       = \frac{(B_{\tilde{\Lambda}} \, A_{\Lambda})_{ij}}{(\lambda_j/\tilde{\lambda}_i)^l-1} 
       \qquad  i,j =1,\ldots,n \, .
\eez
In the scalar case ($m=1$), this leads to $2n$-soliton solutions. The Stein equation is a special case  
of the \emph{Sylvester equation}, about which there is a vast literature. Also solutions with \emph{non}-diagonal 
matrices $\Lambda,\tilde{\Lambda}$ are available. 
\end{remark}

\subsubsection{Exact solutions of the Volterra lattice equation with self-consistent sources}
Let us specialize the results of the preceding subsection to $k=1$ and $l=2$. Then we have\footnote{More 
generally, this satisfies (\ref{V_P,Q-Lambda}) if $l = 2 k$. So these cases can be easily treated as well. }
$\mathfrak{R}(P) = \{ \Lambda, \Lambda^{-1} \}$ and $\mathfrak{R}(Q) = \{ \tilde{\Lambda}, \tilde{\Lambda}^{-1} \}$, 
and thus 
\bez
    \theta = A_1 \, e^{ \Lambda^2 \, t} \, \Lambda^j + A_2 \, e^{ \Lambda^{-2} \, t} \, \Lambda^{-j}
                   \, , \qquad
    \eta = e^{-\tilde{\Lambda}^2 \, t} \, \tilde{\Lambda}^{-j} \, B_1 + e^{-\tilde{\Lambda}^{-2} \, t} \, \tilde{\Lambda}^j \, B_2 \, ,
\eez
with constant matrices $A_a,B_a$, $a=1,2$. Furthermore, 
\bez
    \tilde{\Omega} &=& e^{-\tilde{\Lambda}^2 \, t} \, \tilde{\Lambda}^{-j} \, X_{11} \, e^{ \Lambda^2 \, t} \, \Lambda^j
     + e^{-\tilde{\Lambda}^2 \, t} \, \tilde{\Lambda}^{-j} \, X_{12} \, e^{ \Lambda^{-2} \, t} \, \Lambda^{-j} \\
   &&  + e^{-\tilde{\Lambda}^{-2} \, t} \, \tilde{\Lambda}^j \, X_{21} \, e^{ \Lambda^2 \, t} \, \Lambda^j
     + e^{-\tilde{\Lambda}^{-2} \, t} \, \tilde{\Lambda}^j \, X_{22} \, e^{ \Lambda^{-2} \, t} \, \Lambda^{-j}
       + \tilde{\omega} \, ,
\eez
where the constant matrices $X_{ab}$ have to solve the Stein equations
\bez
  && \tilde{\Lambda}^{-2} \, X_{11} \, \Lambda^2 - X_{11} = B_1 \, A_1 \, , \qquad
     \tilde{\Lambda}^{-2} \, X_{12} \, \Lambda^{-2} - X_{12} = B_1 \, A_2 \, , \\
  && \tilde{\Lambda}^2 \, X_{21} \, \Lambda^2 - X_{21} = B_2 \, A_1 \, , \hspace{1.1cm}
     \tilde{\Lambda}^2 \, X_{22} \, \Lambda^{-2} - X_{22} = B_2 \, A_2 \, .     
\eez
Then
\bez
     g = (I - \theta \, \tilde{\Omega}_{(2)}^{-1} \, \eta) \, g_0  \, , \qquad
     \hat{q} = \theta \, \tilde{\Omega}_{(2)}^{-1} \gamma_i  \, , \qquad
     \tilde{r} = \tilde{\Omega}^{-1} \eta 
\eez
solves (\ref{gVscs1}) (with $k=1$ and $l=2$) if $i=2$ and (\ref{gVscs2}) if $i=1$. The corresponding 
constraint for $\tilde{\omega}$ still has to be satisfied. 
Via (\ref{V_from_g}), this yields solutions of the Volterra lattice equation with self-consistent sources. 
The second type reads
\be
   && V_t - (V \, V_{(1)} - V_{(-1)} \, V) = - ( \hat{q} \, \tilde{r}_{(1)} )_t \, , \nonumber \\
   && \hat{q}_t = V \, V_{(1)} \, \hat{q}_{(2)} \, , \qquad
      \tilde{r}_t = - \tilde{r}_{(-2)} \, V_{(-2)} \, V_{(-1)} \, .  \label{Volterra_with_source}
\ee

\begin{example}
\label{ex:Volterra_2soliton}
 For the scalar ($m=1$) Volterra lattice equation (\ref{Volterra_with_source}), we consider the simplest case, which is $n=1$. 
Here we have constant $\tilde{\omega}$.  
Writing $\Lambda = \sqrt{\lambda}$ and $\tilde{\Lambda} = \sqrt{\tilde{\lambda}}$, 
and assuming $\tilde{\lambda} \notin \{ \lambda, \lambda^{-1} \}$, we obtain
\bez
      g = \lambda \tilde{\lambda} \, g_0 \, \frac{\tau_{(-1)}}{\tau_{(1)}} \, , \qquad
      V = \frac{ \tau_{(-1)} \, \tau_{(2)} }{ \tau \, \tau_{(1)} }  \, ,
\eez
with
\bez
  \tau &=&  A_2 B_1    
        + \frac{\lambda \tilde{\lambda} -1}{\tilde{\lambda} - \lambda} \, [ A_1 B_1 \lambda^j \, e^{(\lambda - \lambda^{-1}) t}
            - A_2 B_2 \, \tilde{\lambda}^j \, e^{(\tilde{\lambda} - \tilde{\lambda}^{-1}) t} ] \\
       &&  - A_1 B_2 \lambda^j \, e^{(\lambda - \lambda^{-1}) t} \, \tilde{\lambda}^j \, e^{(\tilde{\lambda} 
        - \tilde{\lambda}^{-1}) t} 
        + \tilde{\omega}_0 \, (1-\lambda \tilde{\lambda}) \,  (\lambda \tilde{\lambda})^{(j-1)/2} \, 
            e^{( \tilde{\lambda}-\lambda^{-1}) t}     \, .
\eez
In the source-free case ($\tilde{\omega}_0=0$), writing $\lambda = e^\kappa$, $\tilde{\lambda} = e^{\tilde{\kappa}}$,
and choosing the constants $A_i, B_i$ such that all summands are positive, 
the expression for $\tau$ is recognized as the tau function of the 2-soliton solution 
of the scalar version of the Volterra lattice equation (\ref{V_lattice_eq}). This solution 
has previously been obtained via Hirota's bilinear method (see, in particular, \cite{Kaji+Wada98}). 
Fig.~\ref{fig:Volterra_2soliton} shows plots of $V$ for a 2-soliton solution from 
the above family, without and with source, respectively. Fig.~\ref{fig:Volterra_2soliton_qr} displays  
the source term $\hat{q} \, \tilde{r}_{(1)}$. 

\begin{figure}[t] 
\begin{center}
\includegraphics[scale=.2]{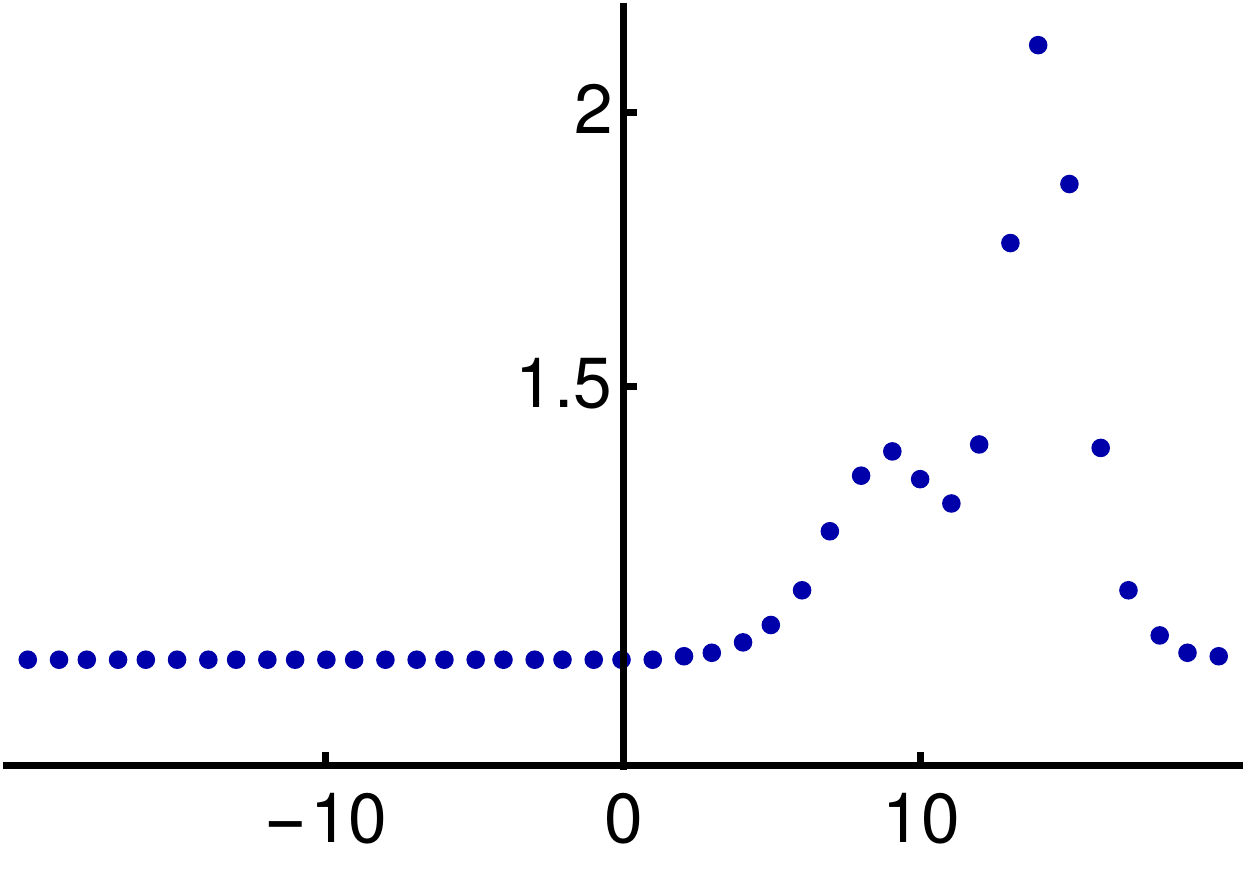}
\includegraphics[scale=.2]{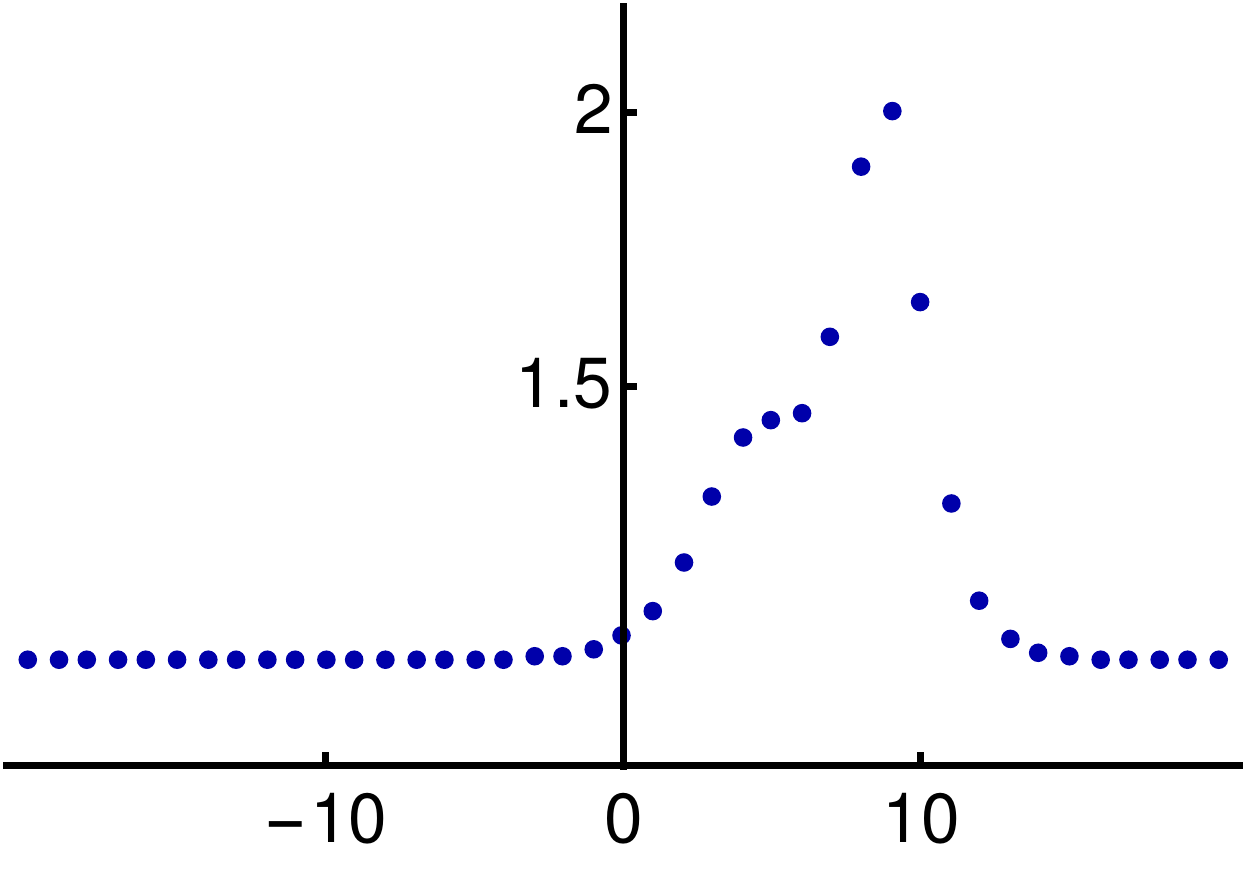}
\includegraphics[scale=.2]{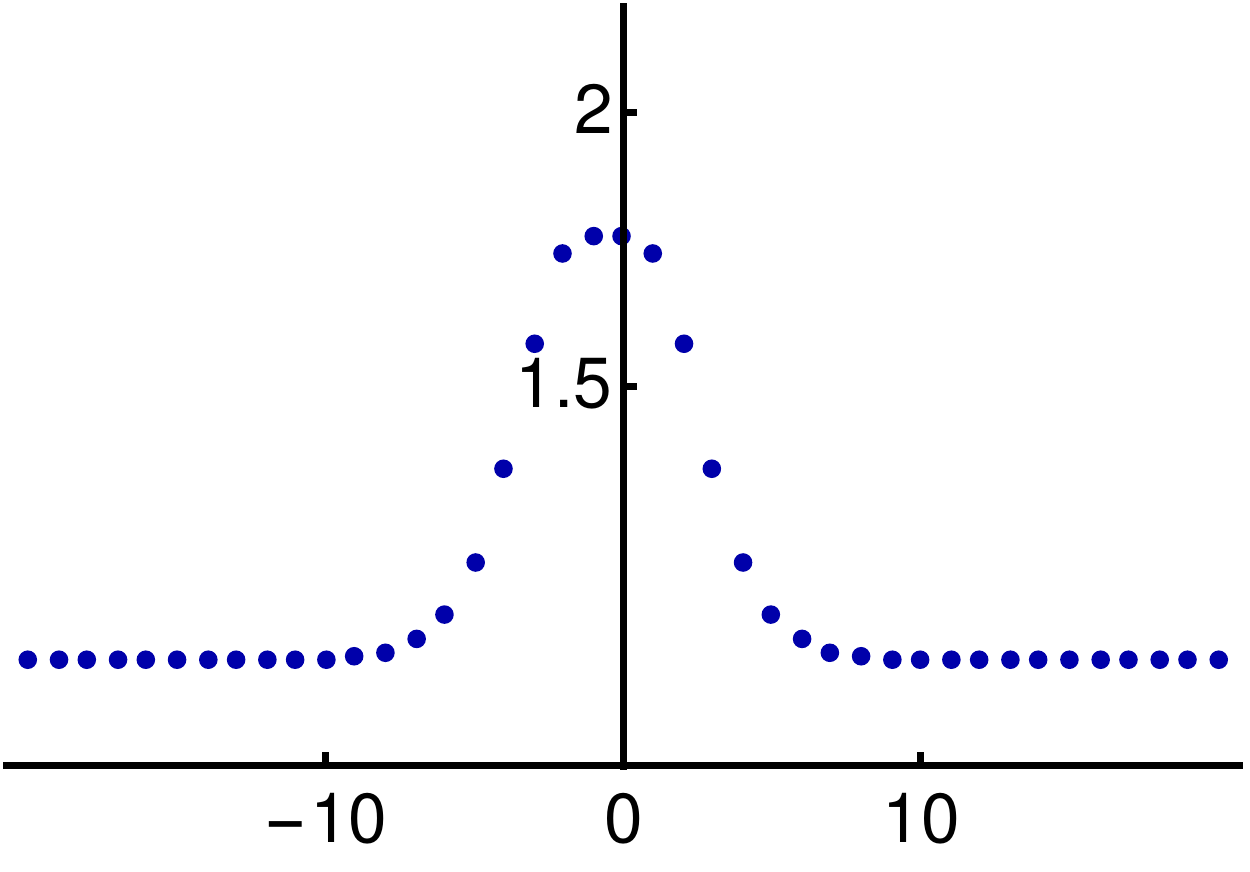}
\includegraphics[scale=.2]{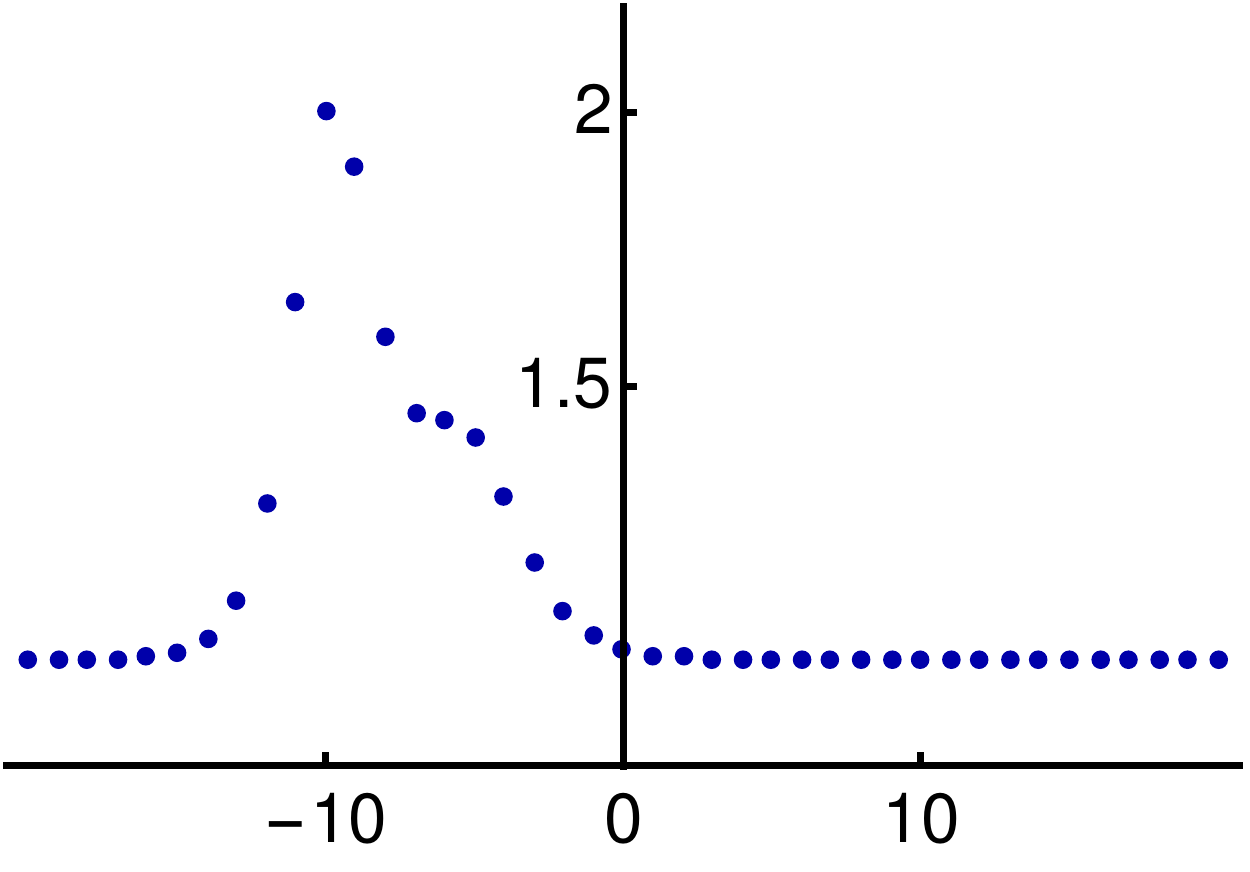}
\includegraphics[scale=.2]{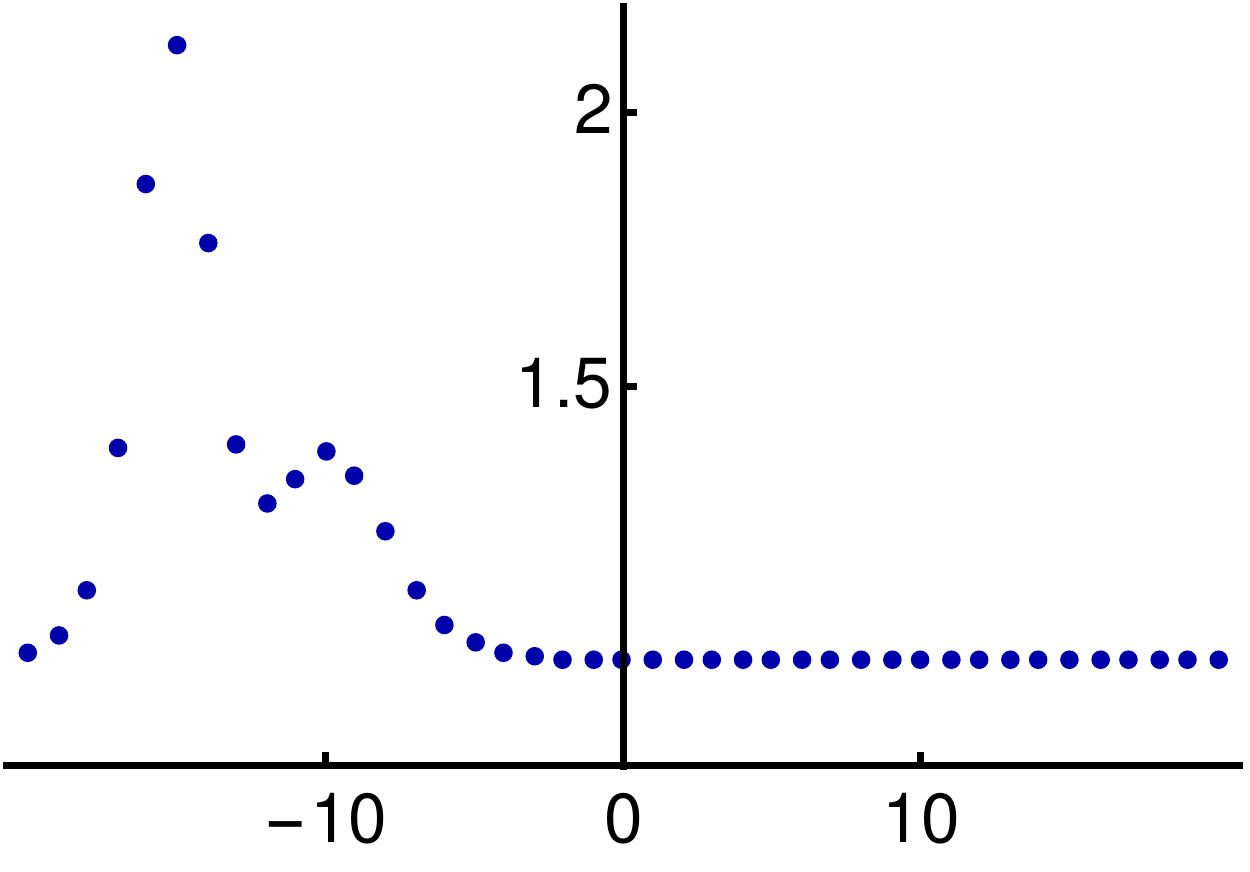}
\vspace{.3cm}

\includegraphics[scale=.2]{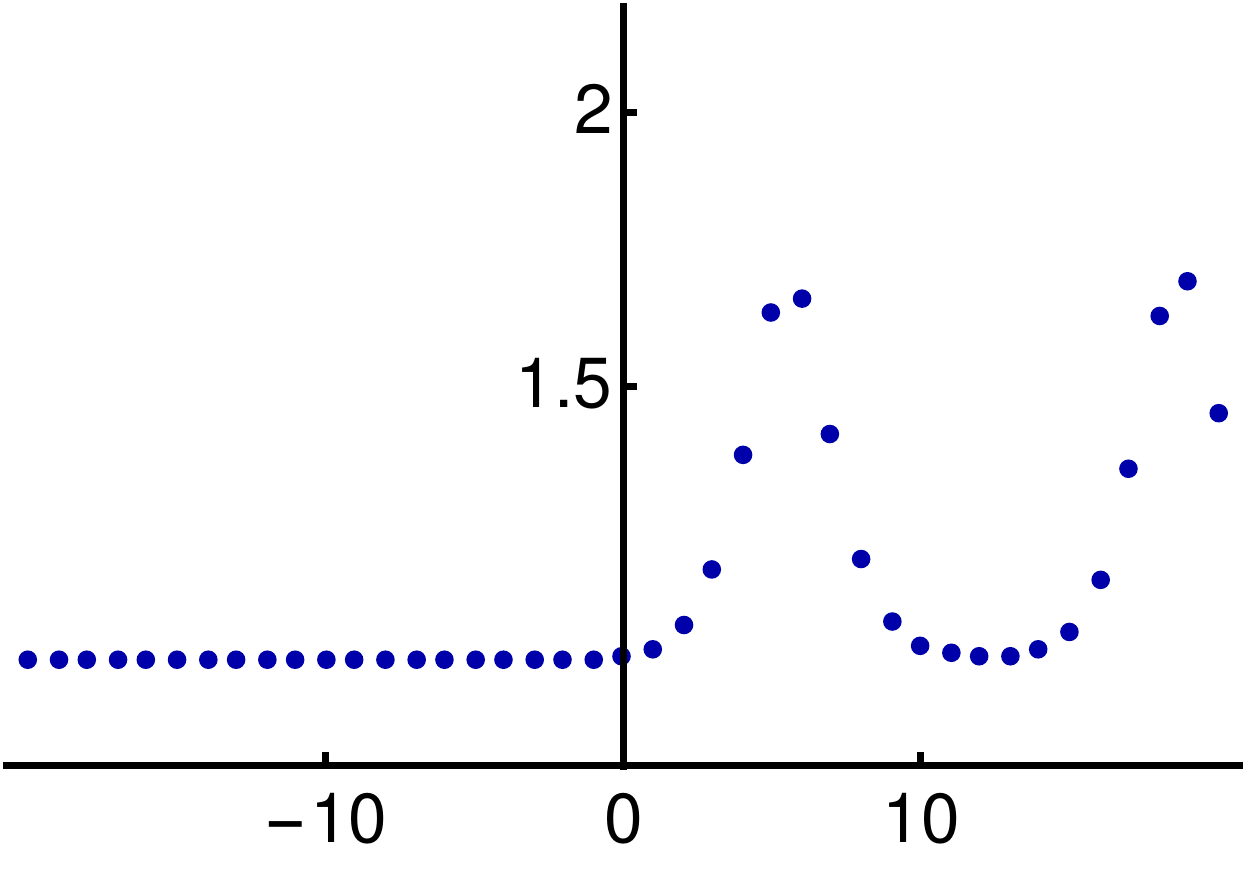}
\includegraphics[scale=.2]{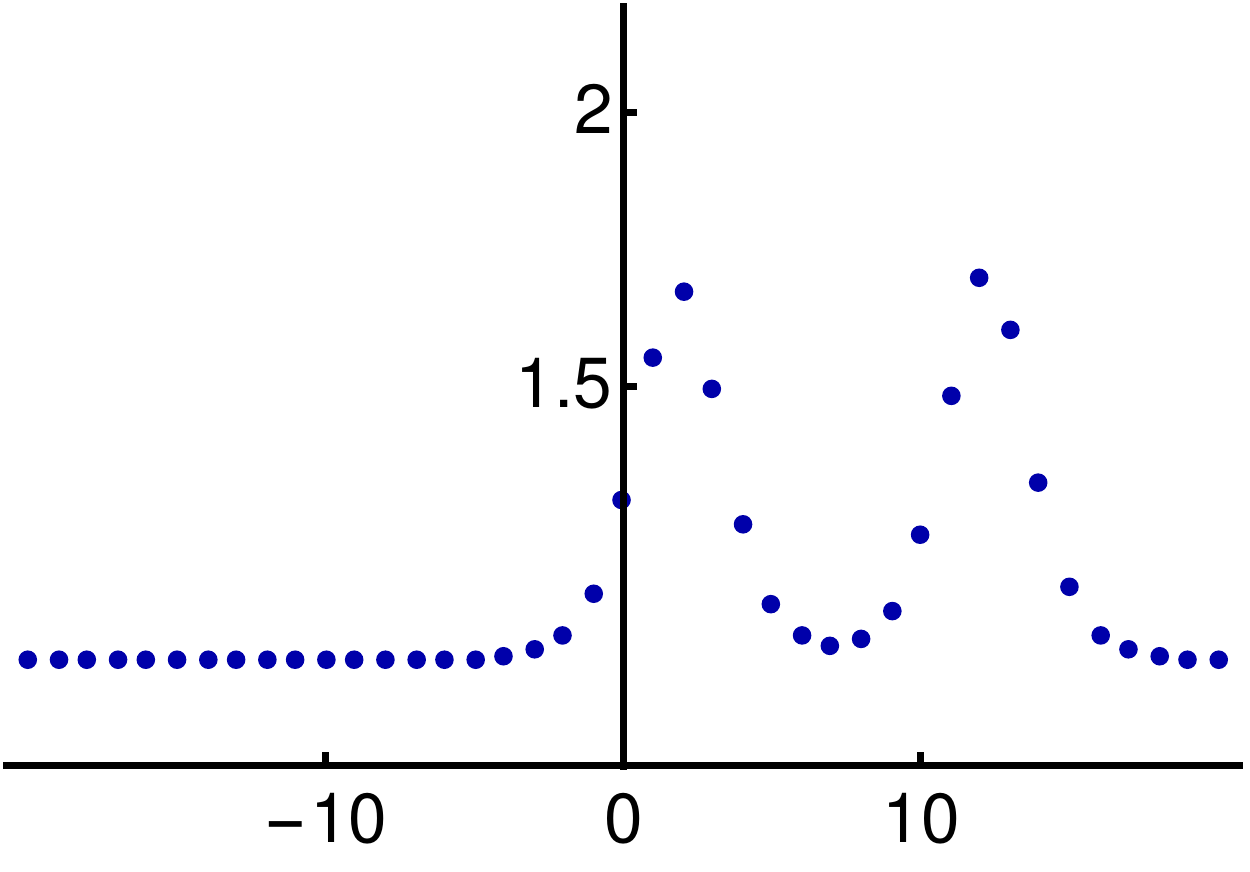}
\includegraphics[scale=.2]{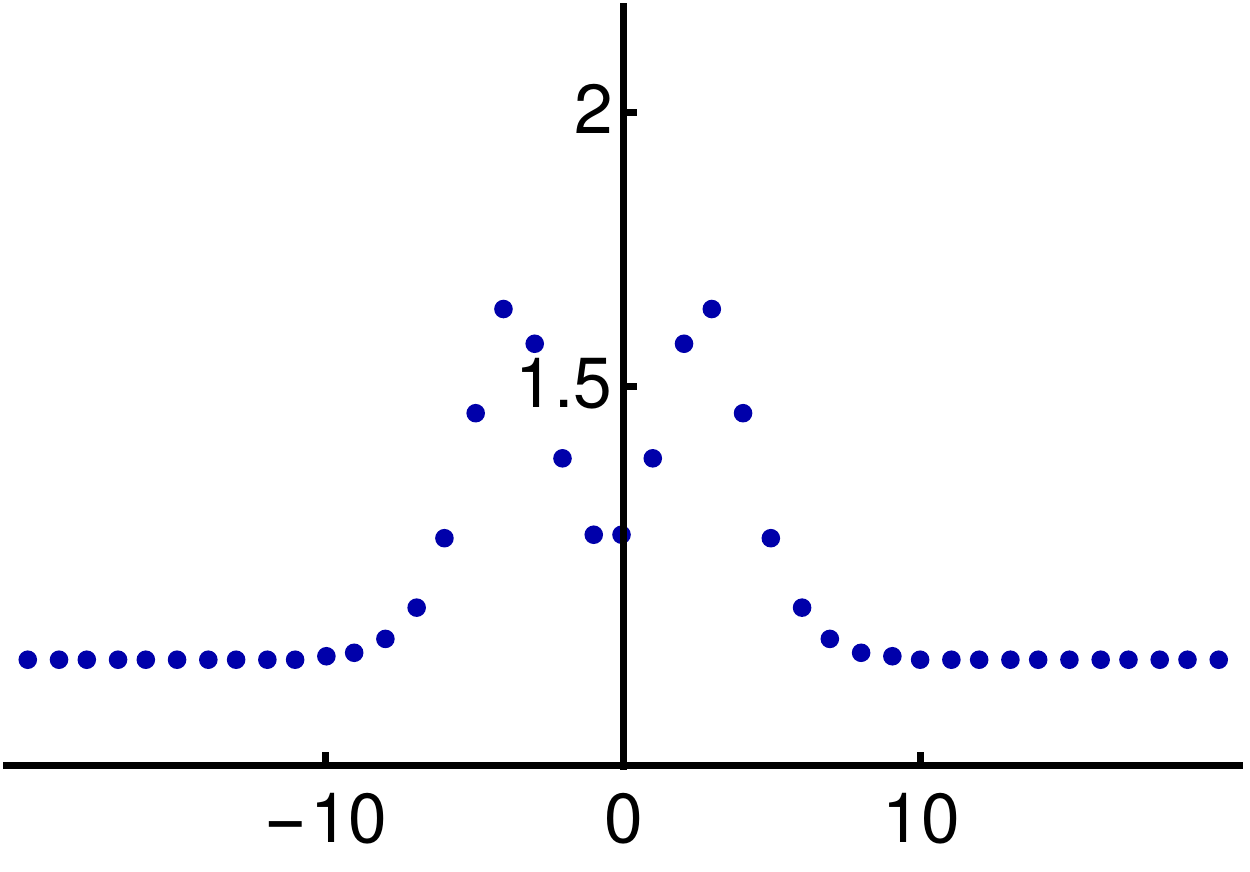}
\includegraphics[scale=.2]{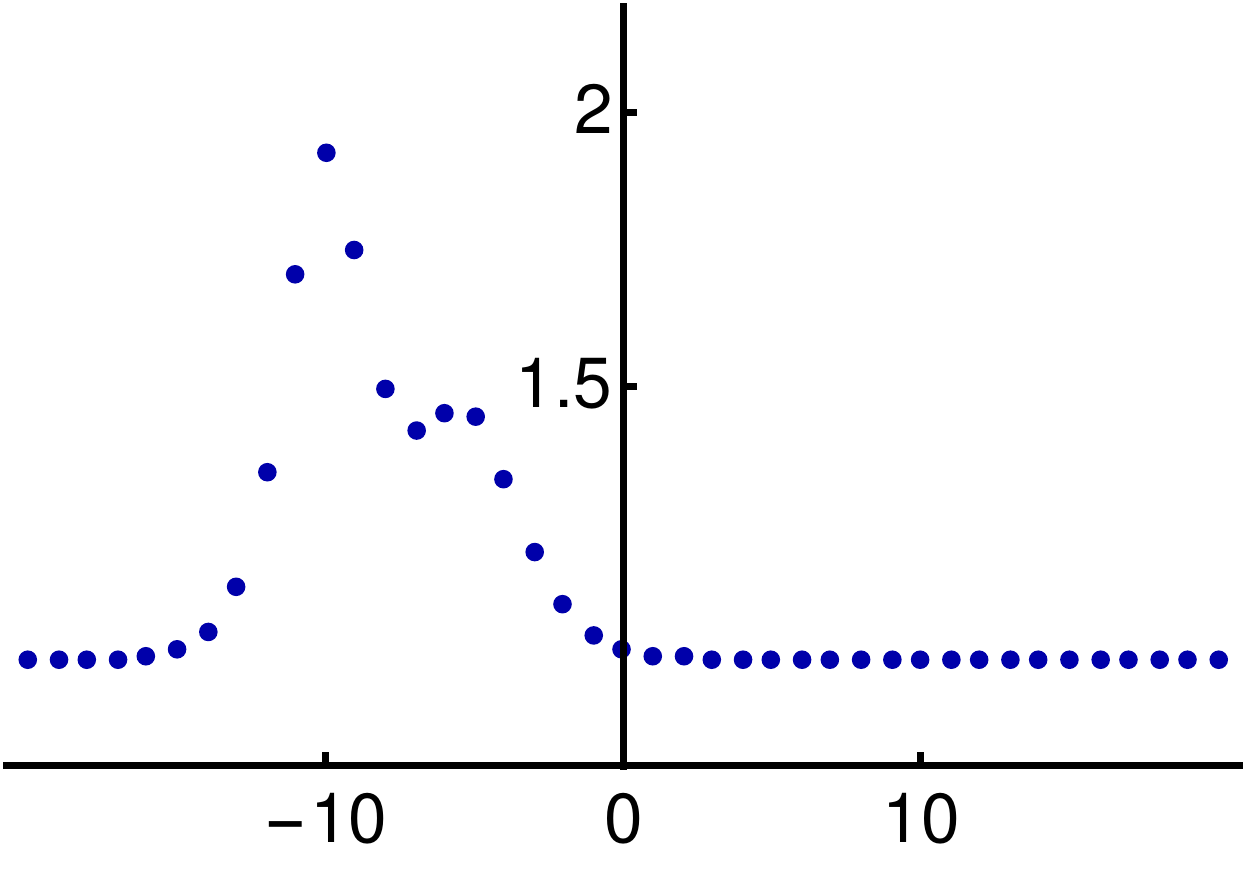}
\includegraphics[scale=.2]{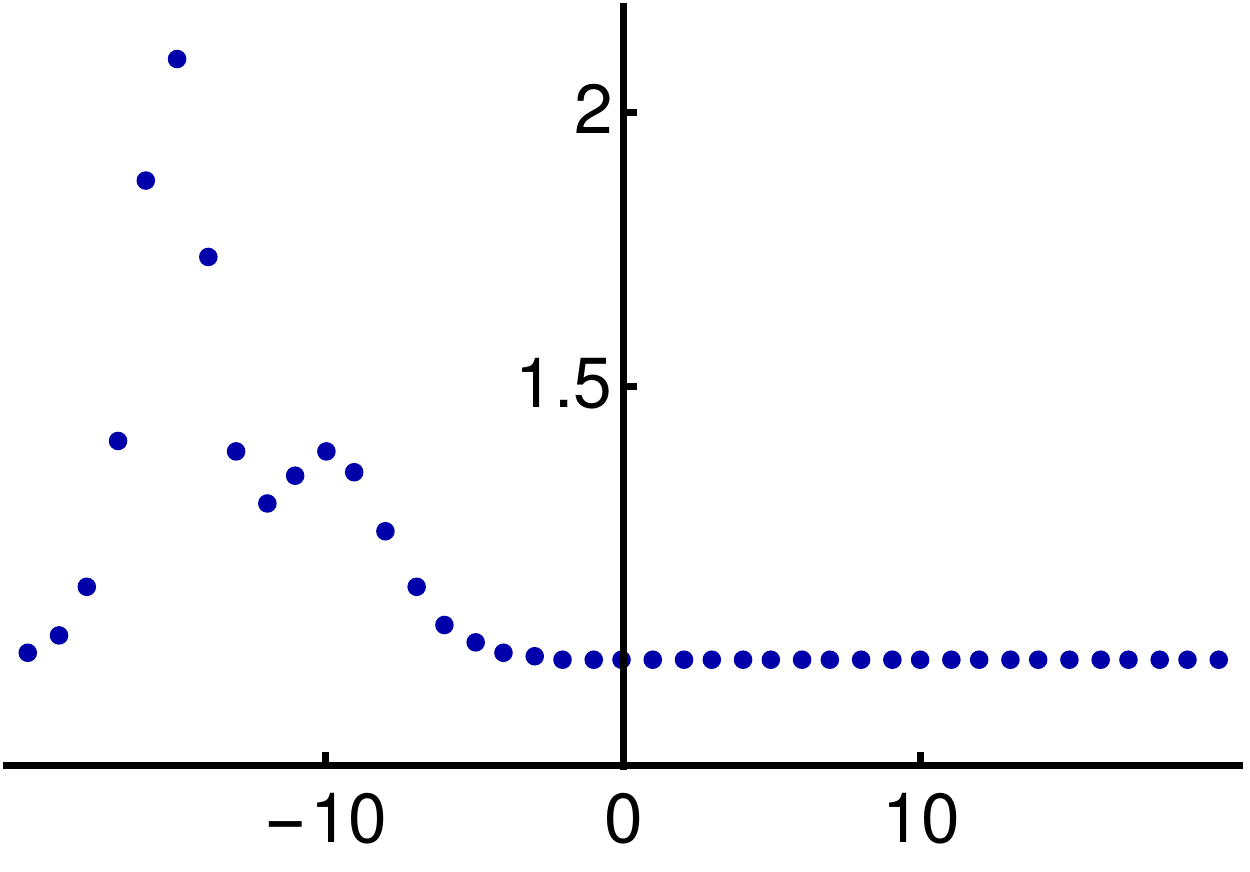}
\end{center}
\caption{Plots of $V$ as a function of the discrete variable $j$, at times $-5,-3,0,3,5$, 
for the 2-soliton solution of the scalar version of the Volterra lattice system (\ref{Volterra_with_source}), 
elaborated in Example~\ref{ex:Volterra_2soliton}. Here we set 
$\lambda=1/2$, $\tilde{\lambda}=3/2$, and $-A_1 = A_2 = B_1 = B_2 =1$. 
The first row shows the solution without sources ($\tilde{\omega}=0$). The second row displays the solution with $\tilde{\omega}=10$. }
\label{fig:Volterra_2soliton} 
\end{figure} 

\begin{figure} 
\begin{center}
\includegraphics[scale=.25]{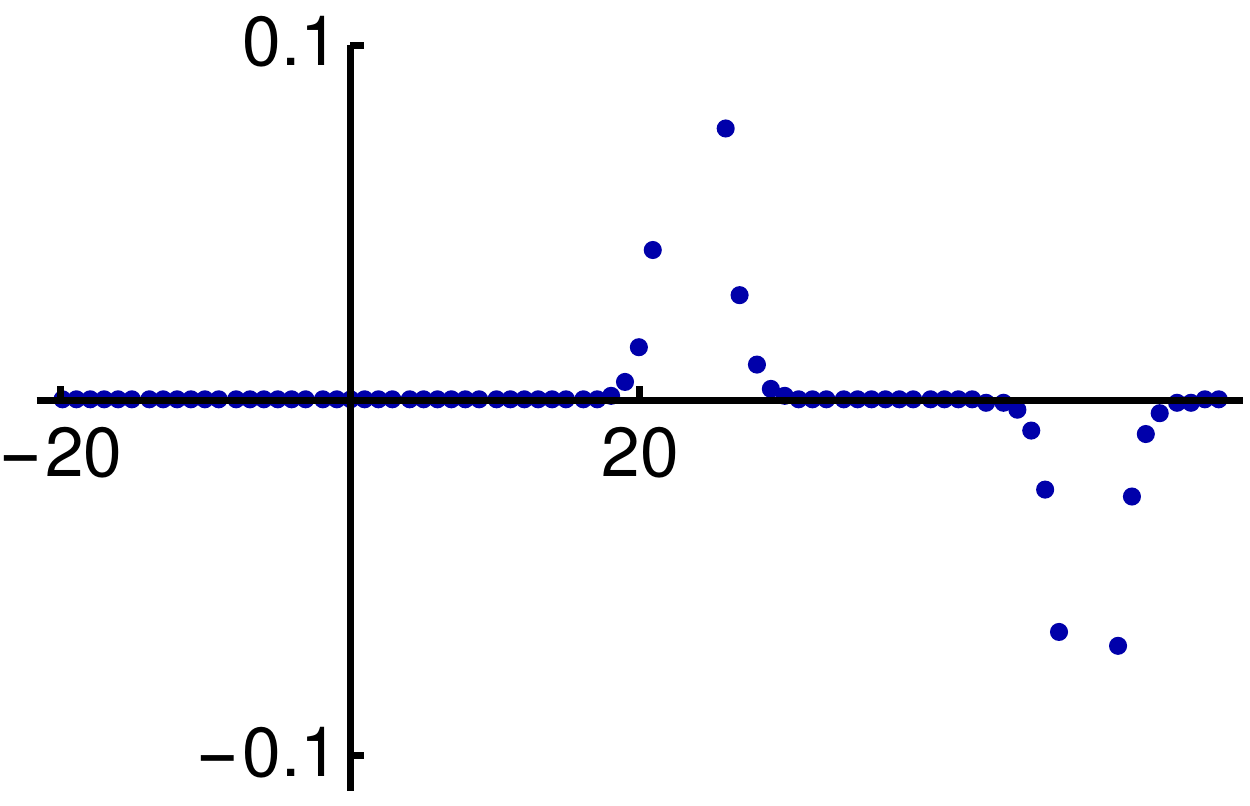}
\hspace{.2cm}
\includegraphics[scale=.25]{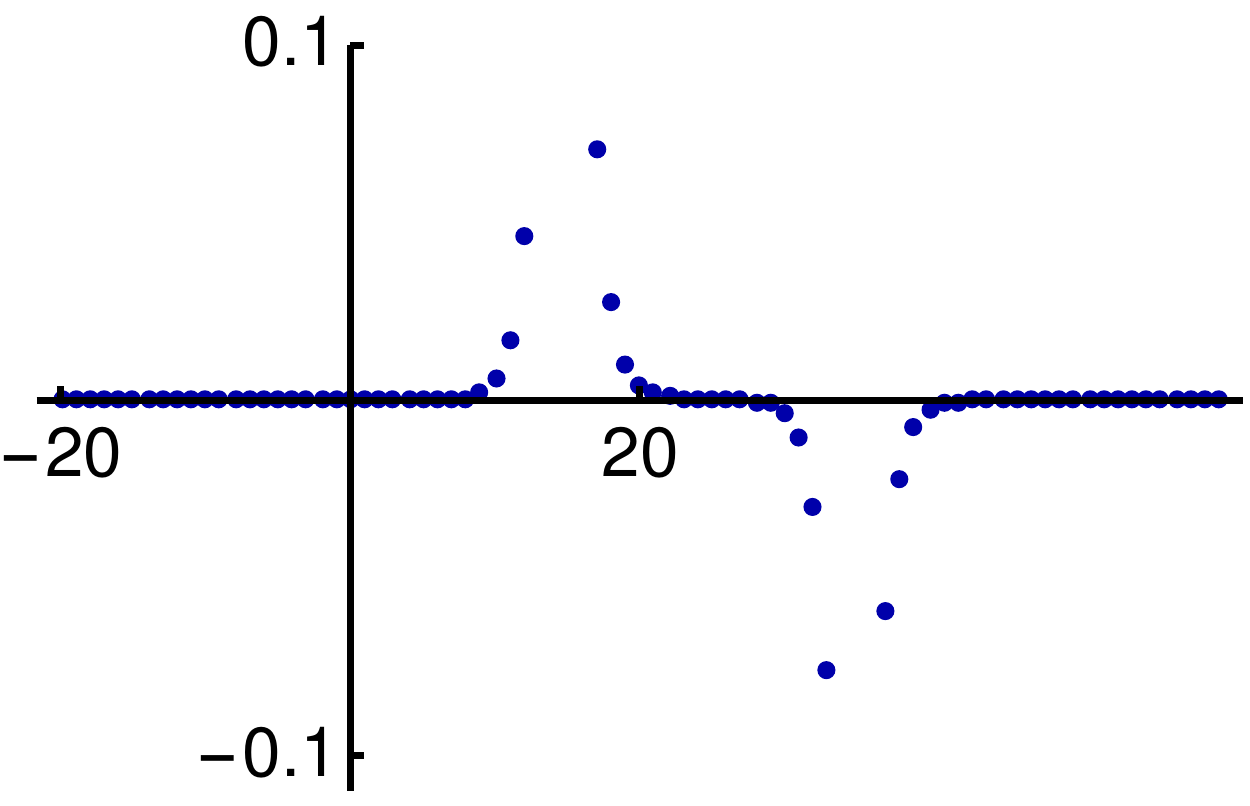}
\hspace{.2cm}
\includegraphics[scale=.25]{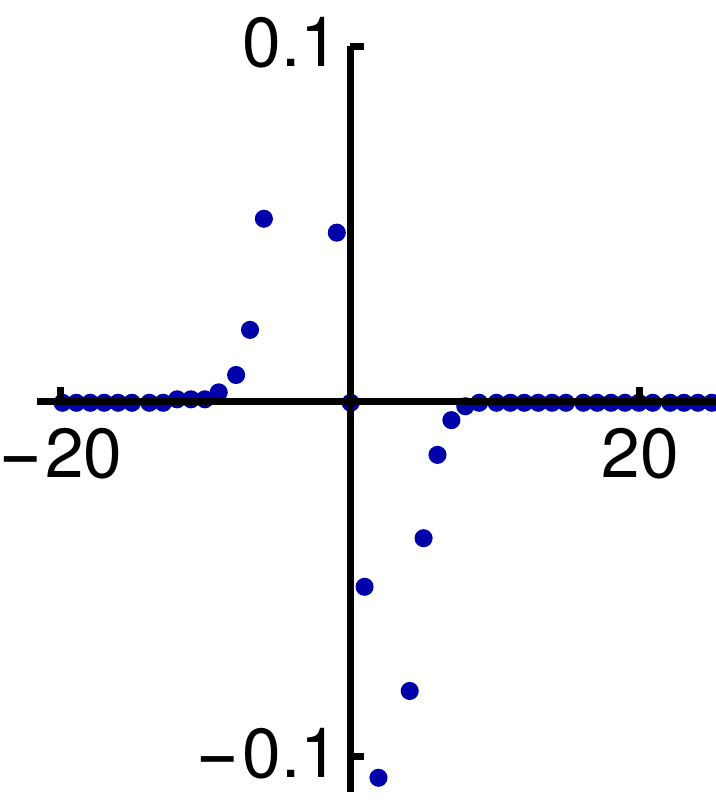}
\hspace{.2cm}
\includegraphics[scale=.25]{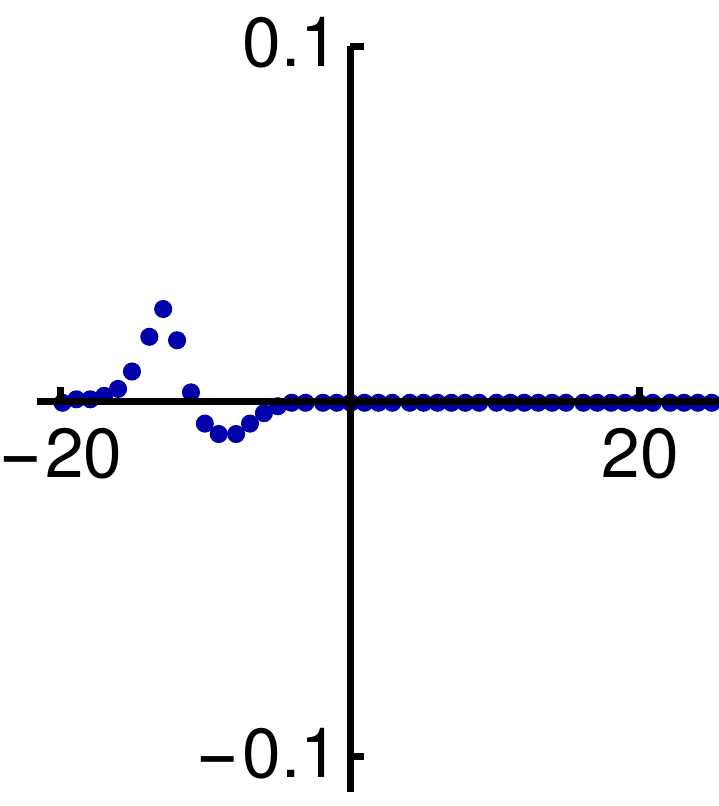}
\end{center}
\caption{Plots of the source term $\hat{q} \, \tilde{r}_{(1)}$ as a function of the discrete variable $j$, 
at times $-15,-10,0,4$, 
for the 2-soliton solution of the scalar Volterra lattice equation with source, for which $V$ has been 
displayed in Fig.~\ref{fig:Volterra_2soliton}. The peaks vanish as $t \to \infty$ and separate into a positive 
and a negative bounded peak as $t \to -\infty$. }
\label{fig:Volterra_2soliton_qr} 
\end{figure}

\end{example}

\subsubsection{Exact solutions of the modified Volterra lattice equation with self-consistent sources}
\label{subsec:mV}
\begin{figure}[t] 
\begin{center}
\includegraphics[scale=.23]{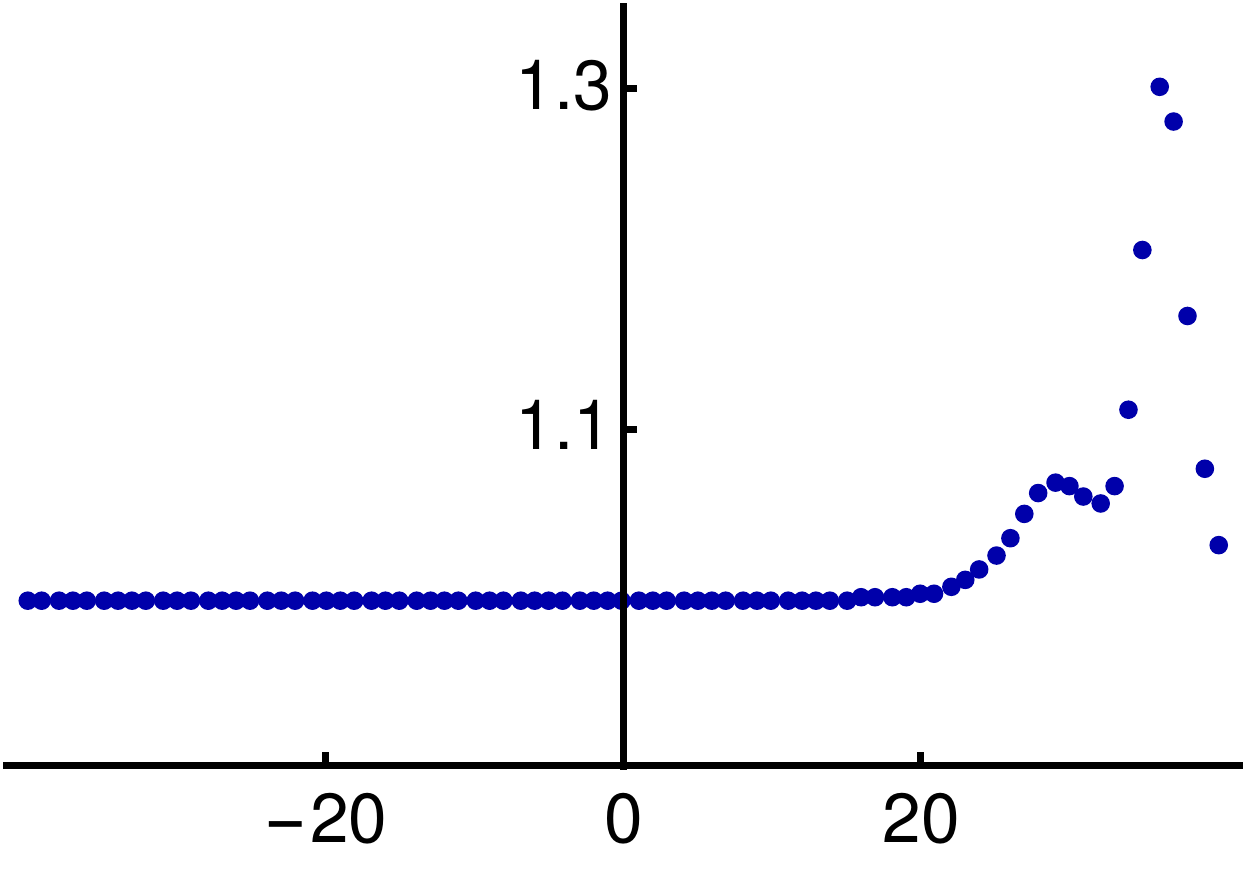}
\includegraphics[scale=.23]{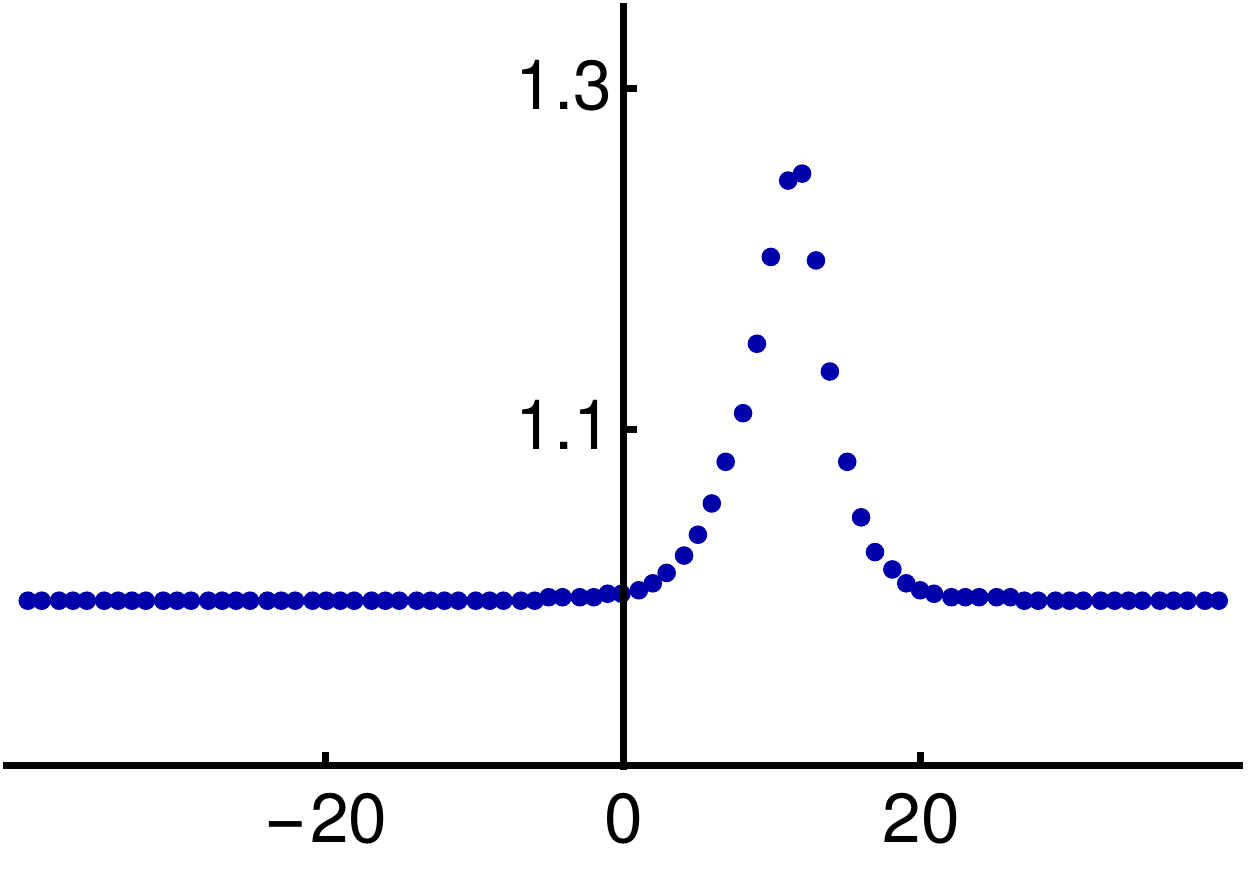}
\includegraphics[scale=.23]{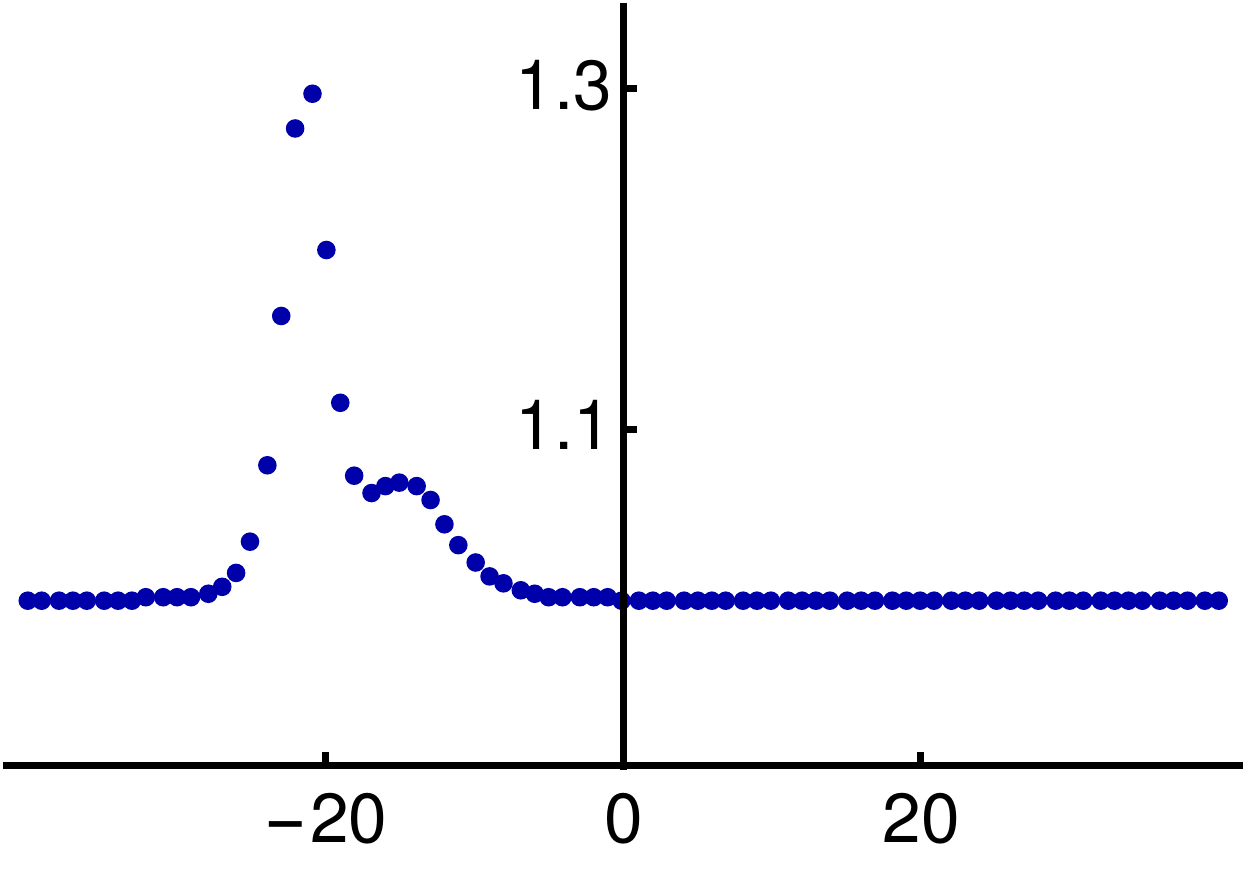}
\includegraphics[scale=.23]{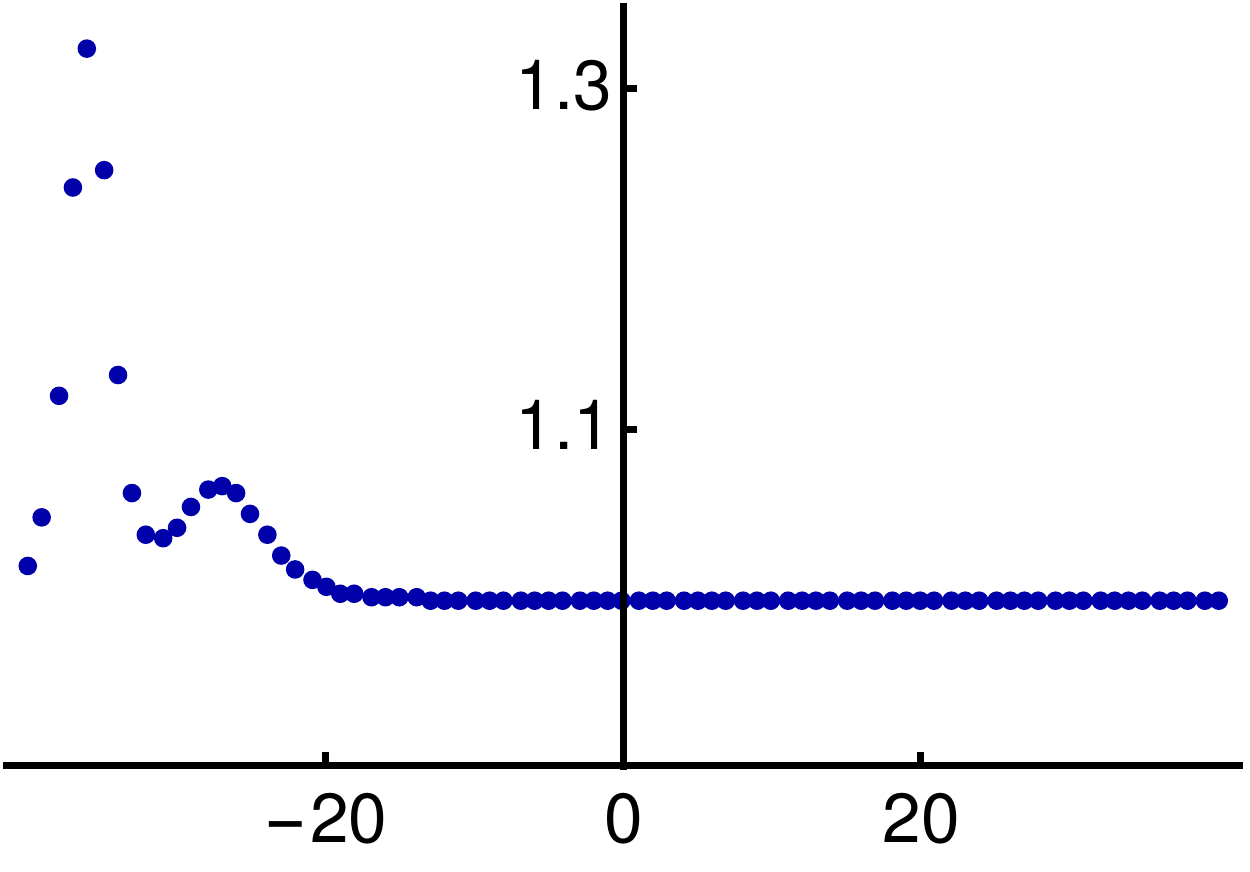} 
\vspace{.3cm}

\includegraphics[scale=.23]{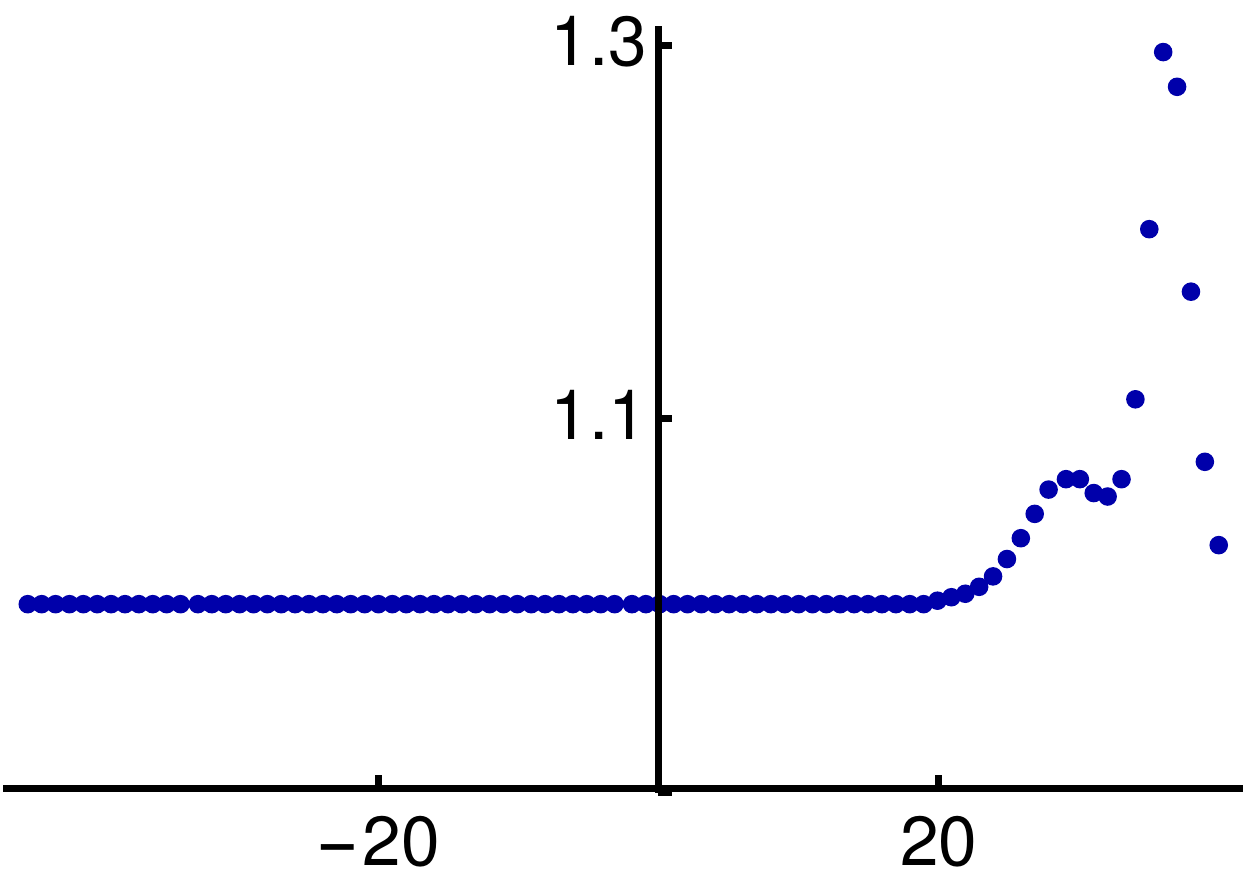}
\includegraphics[scale=.23]{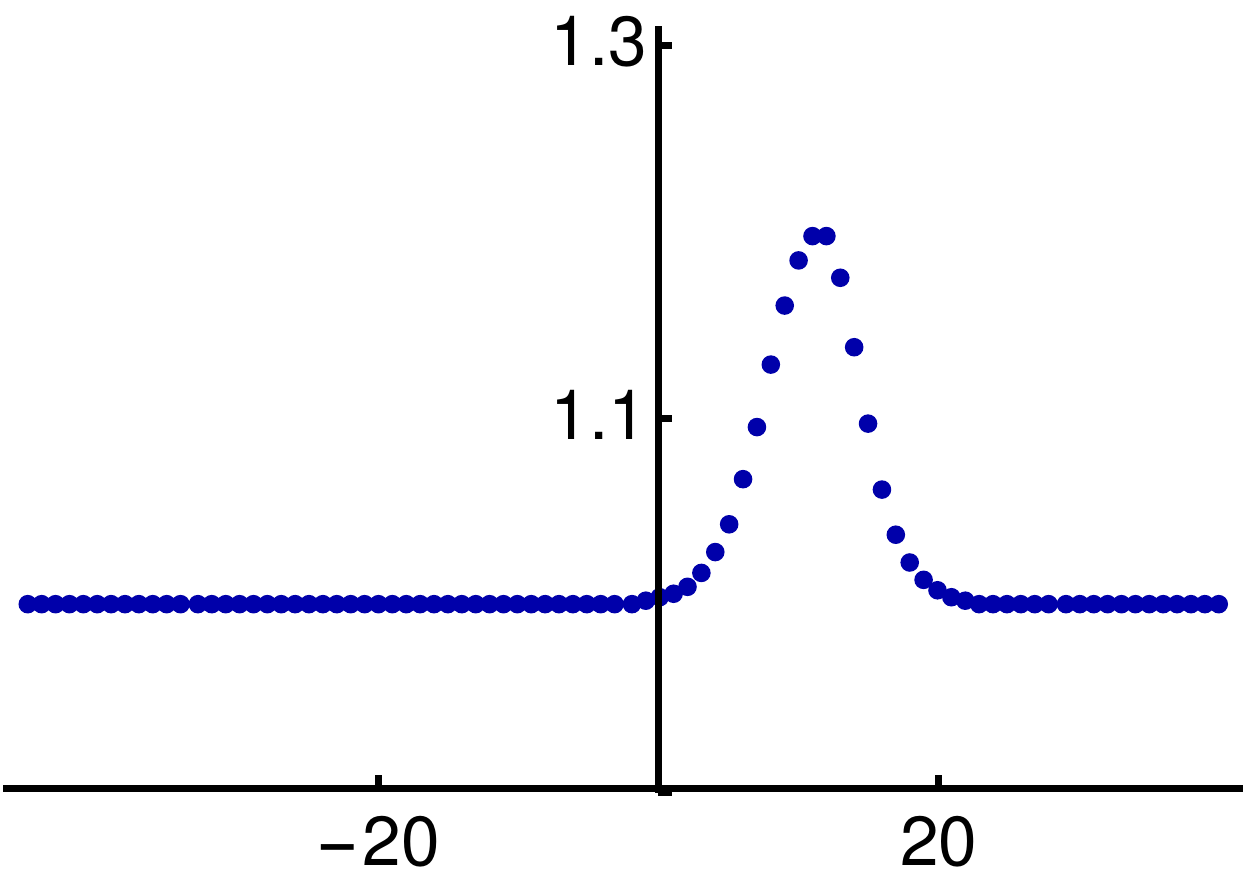}
\includegraphics[scale=.23]{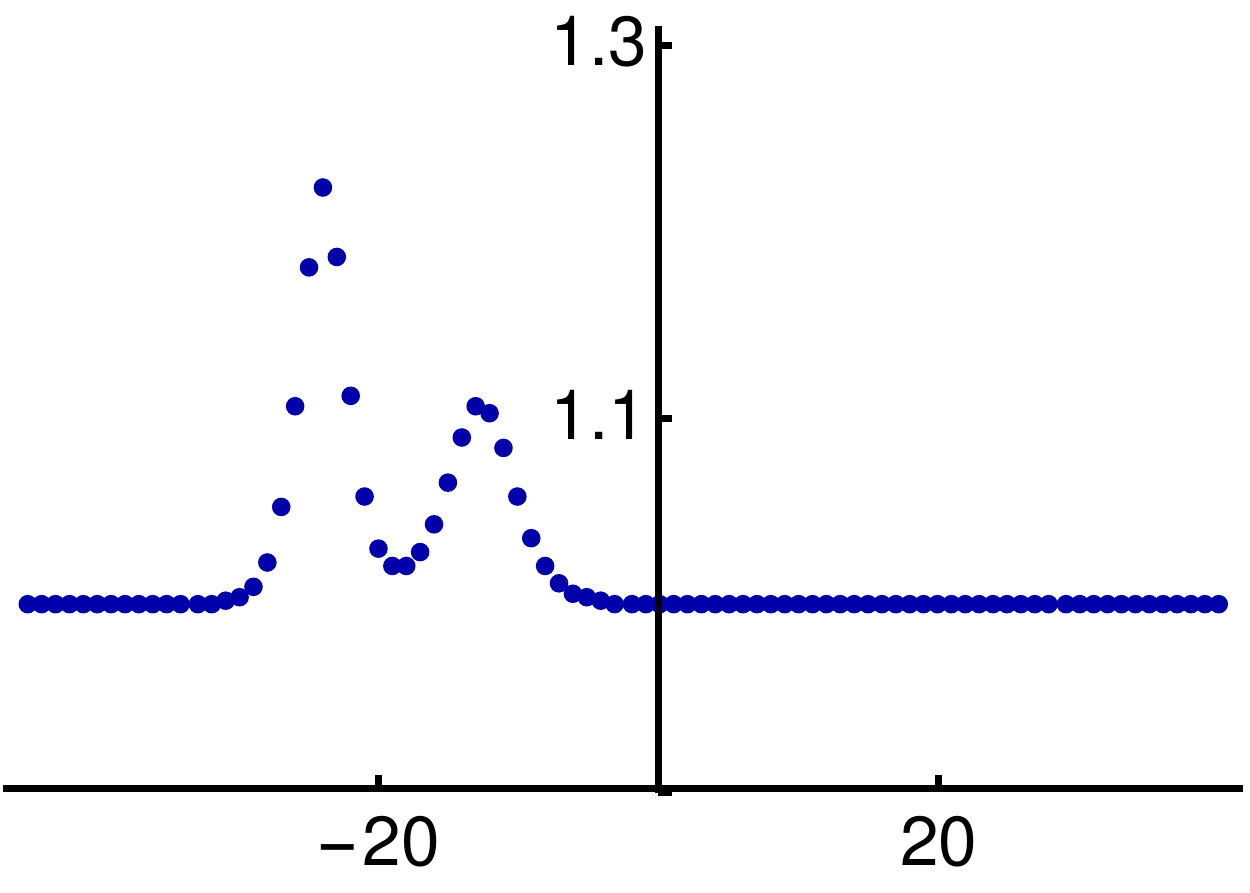}
\includegraphics[scale=.23]{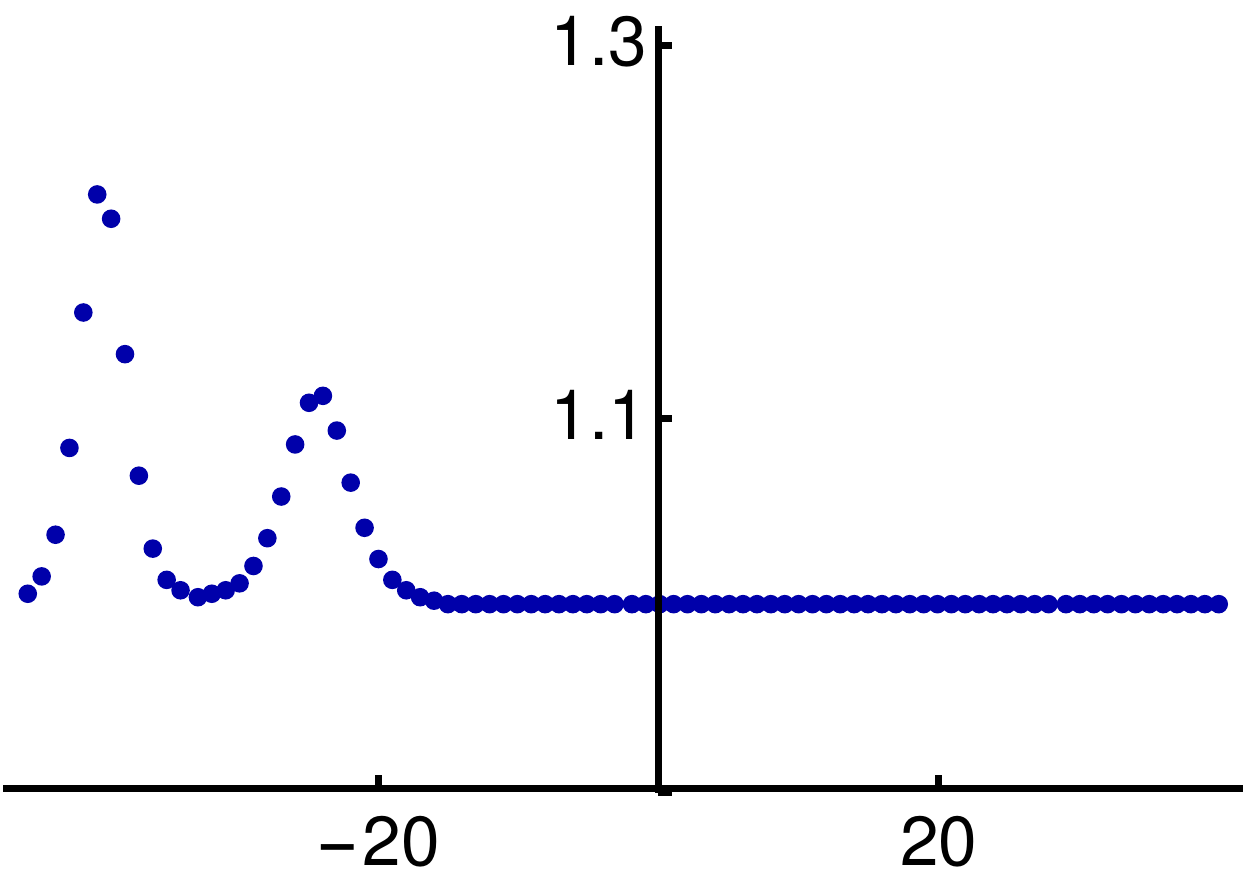}
\end{center}
\caption{Plots of $U$ as a function of the discrete variable $j$, at times $-15,-5,8,14$, 
for the 2-soliton solution of the scalar modified Volterra lattice equation (\ref{mVolterra_with_source}). The first row shows the 
source-free case ($\tilde{\omega}=0$), the second the corresponding solution with $\tilde{\omega} = -10$.
Here we set $M=1/4$, $\tilde{M}=1/8$ and $-A_1 = A_2 = B_1 = B_2 =1$ in Section~\ref{subsec:mV}. }
\label{fig:mVolterra_2soliton} 
\end{figure}

\begin{figure} 
\begin{center}
\includegraphics[scale=.32]{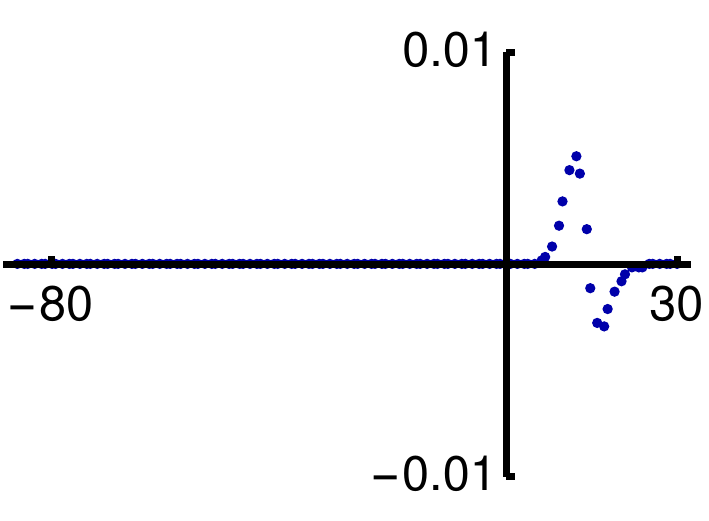}
\hspace{.2cm}
\includegraphics[scale=.32]{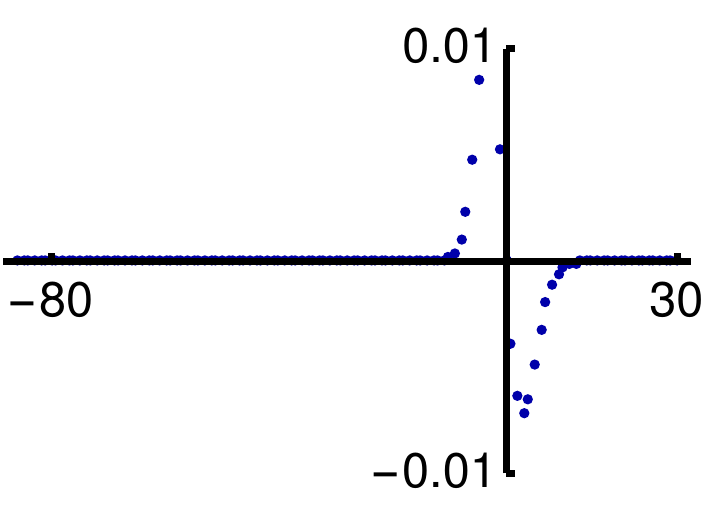}
\hspace{.2cm}
\includegraphics[scale=.32]{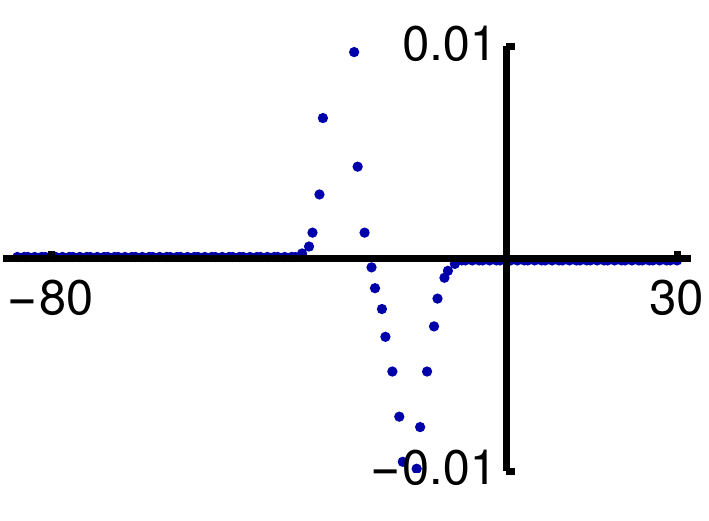}
\hspace{.2cm}
\includegraphics[scale=.32]{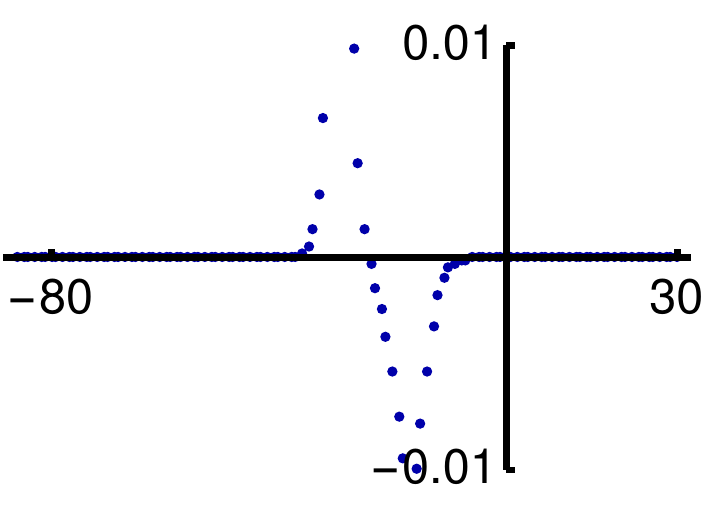}
\hspace{.2cm}
\includegraphics[scale=.32]{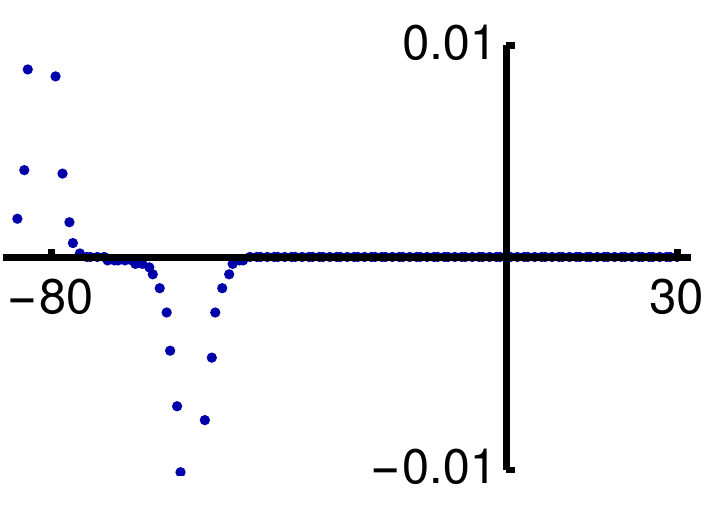}
\end{center}
\caption{Plots of $\hat{q} \, \tilde{r}_{(1)}$ as a function of the discrete variable $j$, at times $-6,0,10,20,30$, 
for the 2-soliton solution of the scalar modified Volterra lattice equation with source, for which $U$ has been 
displayed in Fig.~\ref{fig:mVolterra_2soliton}. The peaks vanish as $t \to -\infty$ and separate into spreading bounded 
peaks as $t$ becomes large. }
\label{fig:mVolterra_2soliton_qr} 
\end{figure}

Here we have to choose $k=1$ and $l=-1$. Setting $P=\frac{1}{4}I-M^2$ and $Q=\frac{1}{4}I-\tilde{M}^2$, 
with arbitrary constant $n \times n$ matrices $M$ and $\tilde{M}$, (\ref{V_P,Q-Lambda}) is satisfied 
by\footnote{For $k=2$ and $l=1$, equations (\ref{V_P,Q-Lambda}) coincide with those for $k=1$ and $l=-1$. 
So this case can be treated analogously. }
\bez
    \Lambda_{1,2} = - \frac{1}{2} I \pm M \, , \qquad
    \tilde{\Lambda}_{1,2} = - \frac{1}{2} I \pm \tilde{M} \, ,
\eez
and we have
\bez
    \theta = A_1 \, e^{ \Lambda_1^{-1} \, t} \, \Lambda_1^j + A_2 \, e^{ \Lambda_2^{-1} \, t} \, \Lambda_2^j
                   \, , \qquad
    \eta = e^{-\tilde{\Lambda}_1^{-1} \, t} \, \tilde{\Lambda}_1^{-j} \, B_1 
           + e^{-\tilde{\Lambda}_2^{-1} \, t} \, \tilde{\Lambda}_2^{-j} \, B_2 \, , 
\eez
with constant matrices $A_a,B_a$, $a=1,2$, and  
\bez
    \tilde{\Omega} &=& e^{-\tilde{\Lambda}_1^{-1} \, t} \, \tilde{\Lambda}_1^{-j} \, X_{11} \, e^{ \Lambda_1^{-1} \, t} \, \Lambda_1^j
     + e^{-\tilde{\Lambda}_1^{-1} \, t} \, \tilde{\Lambda}_1^{-j} \, X_{12} \, e^{ \Lambda_2^{-1} \, t} \, \Lambda_2^j \\
   &&  + e^{-\tilde{\Lambda}_2^{-1} \, t} \, \tilde{\Lambda}_2^{-j} \, X_{21} \, e^{ \Lambda_1^{-1} \, t} \, \Lambda_1^j
     + e^{-\tilde{\Lambda}_2^{-1} \, t} \, \tilde{\Lambda}_2^{-j} \, X_{22} \, e^{ \Lambda_2^{-1} \, t} \, \Lambda_2^j 
       + \tilde{\omega} \, ,
\eez
where the constant matrices $X_{ab}$ have to solve the Stein equations
\bez
  && \tilde{\Lambda}_1 \, X_{11} \, \Lambda_1^{-1} - X_{11} = B_1 \, A_1 \, , \qquad 
     \tilde{\Lambda}_1 \, X_{12} \, \Lambda_2^{-1} - X_{12} = B_1 \, A_2 \, , \\
  && \tilde{\Lambda}_2 \, X_{21} \, \Lambda_1^{-1} - X_{21} = B_2 \, A_1 \, , \qquad
     \tilde{\Lambda}_2 \, X_{22} \, \Lambda_2^{-1} - X_{22} = B_2 \, A_2 \, .     
\eez
Then
\be
     g = (I - \theta \, \tilde{\Omega}_{(-1)}^{-1} \, \eta) \, g_0  \, , \qquad
     \hat{q} = \theta \, \tilde{\Omega}_{(-1)}^{-1} \gamma_i  \, , \qquad
     \tilde{r} = \tilde{\Omega}^{-1} \eta      \label{mV_sol}
\ee
solves (\ref{gVscs1}) (with $k=1$ and $l=-1$) if $i=2$ and (\ref{gVscs2}) if $i=1$, provided that $\tilde{\omega}$ 
satisfies the corresponding constraint. 
Via $U = V^{-1} = g_{(1)} \, g^{-1}$, this yields solutions of the two versions of the \emph{modified} Volterra lattice 
equation with self-consistent sources.
If $i=1$, in which case $\tilde{\omega}$ is constant, the latter reads (cf. (\ref{gmVolterra_eqs}))
\be
  &&  U_t - U \, (U_{(1)} - U_{(-1)}) \, U = U \, ( \hat{q} \, \tilde{r}_{(1)} )_t \, U \, , \nonumber \\
  && \hat{q}_t = U_{(-1)} \, \hat{q}_{(-1)} \, , \qquad  
     \tilde{r}_t = - \tilde{r}_{(1)} \, U  \, .    \label{mVolterra_with_source}
\ee
Fig.~\ref{fig:mVolterra_2soliton} shows plots of a 2-soliton solution of the scalar version of this system ($m=n=1$). 
Fig.~\ref{fig:mVolterra_2soliton_qr} displays the source term $\hat{q} \, \tilde{r}_{(1)}$. 

\section{On the second family of generalized Volterra lattice equations}
\label{sec:Volterra2}
The equation (\ref{g_eq}) in bidifferential calculus possesses a ``Miura-dual'' \cite{DMH08bidiff}, which is
\be
    \d \bd \phi + \d \phi \, \d \phi = 0 \,      \label{phi_eq}
\ee
with $\phi \in \mathrm{Mat}(m,m,\mathcal{A})$. For this equation, there is 
the following counterpart of Theorem~\ref{thm:main_g} \cite{CDMH16}.

\begin{theorem}
\label{thm:main_phi}
Let $\phi_0$ be a solution of (\ref{phi_eq}) and let $\theta$, $\eta$, $\Omega$ be solutions of the linear equations 
\be
  &&  \bd \theta = (\d \phi_0) \, \theta + (\d \theta) \, \Delta + \theta \, \alpha \, ,  \label{phi_theta_eq}     \\    
  &&  \bd \eta = - \eta \, \d \phi_0 + \Gamma \, \d \eta + \beta \, \eta \, ,   \label{phi_eta_eq} \\
  &&  \Gamma \, \Omega - \Omega \, \Delta = \eta \, \theta \, ,   \label{Omega_Sylvester} \\
  &&  \bd \Omega = (\d \Omega) \, \Delta - (\d \Gamma) \, \Omega 
     + \beta \, \Omega + \Omega \, \alpha + (\d \eta) \, \theta + \gamma \, ,  \label{bd_Omega} 
\ee
where $\Delta, \Gamma, \alpha, \beta$ satisfy (\ref{Delta,alpha,Gamma,beta_eqs}), $\gamma$ is given by 
(\ref{gamma_in_terms_of_omega}) in terms of $\omega$, which has to satisfy (\ref{c=0_constraint}). Then
\be
    \phi = \phi_0 - \theta \, \Omega^{-1} \, \eta \, , \qquad
       q = \theta \, \Omega^{-1} \, , \qquad 
       r = \Omega^{-1} \, \eta    \label{phi,q,r}
\ee
constitutes a solution of
\be
    \d \, \bd \, \phi + \d \phi \; \d \phi = \d ( q \, \gamma \, r )  \label{scs_phi_eq}
\ee
and 
\be
  &&  \bd q = (\d \phi) \, q + \d( q \, \Gamma) - q \, \beta - q \, \gamma \, \Omega^{-1}  \, , \nonumber \\
  &&  \bd r = - r \, \d \phi + \d( \Delta \, r) - \alpha \, r  - \Omega^{-1} \, \gamma \, r  \, .  \label{scs_q,r_eqs}
\ee
\hfill $\Box$
\end{theorem}
\vspace{.2cm}

If a partial differential-difference equation can be realized via (\ref{phi_eq}), by specifying the 
bidifferential calculus, then Theorem~\ref{thm:main_phi} allows to derive a corresponding binary 
Darboux transformation and also self-consistent source extensions (if $\gamma \neq 0$). 

\begin{remark}
Exploiting (\ref{phi_eq}), using the bidifferential calculus determined by (\ref{sdcm_bidiff}) and with $\phi = \varphi \, \bbS_2$, 
leads to 
\bez
   (\varphi_{,1} - \varphi)_t + (\varphi_t - I) (\varphi_{,2}-\varphi_{,1}) - (\varphi_{,2}-\varphi_{,1})_{,-2} (\varphi_t - I)_{,1} = 0 \, .
\eez
Using (\ref{sdcm_redefinitions}), so that (\ref{P,Q_etc_eqs}), (\ref{sdcm_c=0}) and (\ref{sdcm_gamma}) holds, one can derive 
corresponding self-consistent source extensions, a binary Darboux transformation and exact solutions. Via (\ref{V_shifts}) the 
above equation leads to 
\bez
    (\tilde{\varphi}_{(k)} - \tilde{\varphi})_t + \tilde{\varphi}_t \, (\tilde{\varphi}_{(l)}-\tilde{\varphi}_{(k)}) 
              - (\tilde{\varphi}_{(l)}-\tilde{\varphi}_{(k)})_{(-l)} \, \tilde{\varphi}_{(k) t} = 0 \, ,  
\eez
where $\tilde{\varphi} = \varphi - t \, I$.
But this equation is not (directly) related to any of the known families of generalized Volterra lattices. 
\end{remark}

In the following, we exploit Theorem~\ref{thm:main_phi} using the bidifferential calculus determined by (\ref{sdcm_bidiff}),
but with the roles of $\d$ and $\bd$ exchanged. Setting 
\bez
        \phi = \varphi \, \bbS_2^{-1} \, ,
\eez
(\ref{phi_eq}) then leads to
\be
   (\varphi_{,1} - \varphi)_t = (\varphi_{,2} - \varphi)(\varphi_{,1} - \varphi +I)
   -   (\varphi_{,1} - \varphi +I) ( \varphi_{,2} - \varphi)_{,1,-2} \, .   \label{2+1_varphi_eq}
\ee
An equation equivalent to (\ref{2+1_varphi_eq}) appeared in \cite{Bogo91} (see the theorem on page 41 therein). 
Applying the reduction (\ref{V_shifts}), we obtain (\ref{varphi_preW_eqs}). Via (\ref{W_from_varphi}) 
this results in the second family (\ref{genW_eqs}) of generalized Volterra lattice equations. 

\begin{remark} 
According to Remark~\ref{rem:d,bd_exchange_for_g-eq}, we could also have addressed the first family of generalized 
Volterra lattice equations using (\ref{sdcm_bidiff}) with the expressions for $\d$ and $\bd$ exchanged, thus treating 
both families using the same bidifferential calculus. Though this is indeed true, the elaboration of the first family 
would then have required a little more effort than what was needed in Section~\ref{sec:Volterra1}. 
\end{remark}

\subsection{Self-consistent source extensions of (\ref{2+1_varphi_eq})} 
\label{subsec:scs_phi}
We set
\bez
  && q = \tilde{q} \, \bbS_2^{-1} \, , \quad
     r = \bbS_2^{-1} \, \tilde{r} \, , \quad
     \Delta = \Gamma = \bbS_2^{-1} \, , \\
  &&  \alpha = (P + 2 I) \, \bbS_1 \bbS_2^{-1} \, \xi_1 + \xi_2 \, , \quad
      \beta = - (Q + 2 I) \, \bbS_1 \bbS_2^{-1} \, \xi_1 - \xi_2  \, , \quad  \\
  && \Omega = - \bbS_2  \, \tilde{\Omega} \,  , \quad 
     \omega = \tilde{\omega} \, \bbS_2 \, , \quad 
     \gamma = \gamma_1 \, \bbS_1 \, \xi_1 + \gamma_2 \, \bbS_2 \, \xi_2   \, . 
\eez
Then $\tilde{q}, \tilde{r}, P, Q, \tilde{\omega}$ and $\tilde{\Omega}$ can be restricted to be matrices over 
$\cA_0$ instead of $\cA$ (with the algebras chosen in Section~\ref{sec:Volterra1}). 
(\ref{Delta,alpha,Gamma,beta_eqs}) leads to (\ref{P,Q_etc_eqs}), (\ref{c=0_constraint}) to (\ref{sdcm_c=0}), 
and $\gamma_1$ and $\gamma_2$ are again given by the expressions (\ref{sdcm_gamma}). We will assume that 
$P$ and $Q$ are constant and that $\tilde{\omega}$ only depends on $t$.

(\ref{scs_phi_eq}) reads 
\be
 &&  (\varphi_{,1} - \varphi)_t 
    + (\varphi_{,1} - \varphi + I) ( \varphi_{,2} - \varphi)_{,1,-2} - (\varphi_{,2} - \varphi)(\varphi_{,1} - \varphi + I)
       \hspace{1cm}   \nonumber \\
 && \hspace{1.cm} =  ( \tilde{q}_{,2} \, \gamma_1 \,\tilde{r}_{,1,-2})_{,-2} - \tilde{q}_{,2} \, \gamma_1 \,\tilde{r}_{,1,-2}
    + ( \tilde{q} \, \gamma_2 \, \tilde{r}_{,-2} )_{,1} - \tilde{q} \, \gamma_2 \, \tilde{r}_{,-2} \, .
         \qquad       \label{2+1_varphi_eq_scs}
\ee
(\ref{phi_theta_eq}) and (\ref{phi_eta_eq}) take the form
\be
    \tilde{q}_{,1,2} &=& - (\varphi_{,1} - \varphi + I)_{,2} \, \tilde{q}_{,1} - \tilde{q}_{,2} \, Q
                       - \tilde{q}_{,2} \, \gamma_1 \, \tilde{\Omega}_{,1}^{-1} \, , \nonumber \\
    \tilde{q}_t &=& \tilde{q}_{,2} + (\varphi_{,2} - \varphi) \, \tilde{q} 
                    + \tilde{q} \gamma_2 \, \tilde{\Omega}^{-1} \, , \nonumber \\
    \tilde{r}_{,-2} &=& - \tilde{r} \, (\varphi_{,1} - \varphi + I) - P \, \tilde{r}_{,1,-2} 
                        + \tilde{\Omega}_{,2}^{-1} \, \gamma_1 \, \tilde{r}_{,1,-2} \, , \nonumber \\
    \tilde{r}_t &=&  -\tilde{r}_{,-2} - \tilde{r} \, (\varphi_{,2} - \varphi) 
                     + \tilde{\Omega}_{,2}^{-1} \, \gamma_2 \, \tilde{r}  \, .   \label{2+1_q,r_eqs}
\ee
We obtain the following self-consistent source extensions of (\ref{2+1_varphi_eq}).

\begin{enumerate}
\item $\gamma_1 =0$, i.e., $Q \, \tilde{\omega} = \tilde{\omega} \, P$. In terms of 
$\hat{q} = \tilde{q} \, \gamma_2$ we obtain
\bez
 && (\varphi_{,1} - \varphi)_t + (\varphi_{,1} - \varphi +I) ( \varphi_{,2} - \varphi)_{,1,-2}
    - (\varphi_{,2} - \varphi)(\varphi_{,1} - \varphi +I)  \\
 && \hspace{.75cm}  = ( \hat{q} \, \tilde{r}_{,-2} )_{,1} - \hat{q} \, \tilde{r}_{,-2} \, ,  \\
 &&  \hat{q}_{,1,2} = -(\varphi_{,1} - \varphi +I)_{,2} \, \hat{q}_{,1} - \hat{q}_{,2} \, P \, , \\
 &&  \tilde{r}_{,-2} = - \tilde{r} \, (\varphi_{,1} - \varphi +I) - P \, \tilde{r}_{,1,-2}  \, . 
\eez
\item $\gamma_2 =0$, i.e., $\tilde{\omega}_t = 0$. We introduce $\hat{q} = \tilde{q} \, \gamma_1$. 
\bez
 && (\varphi_{,1} - \varphi)_t + (\varphi_{,1} - \varphi +I) ( \varphi_{,2} - \varphi)_{,1,-2} 
    - (\varphi_{,2} - \varphi)(\varphi_{,1} - \varphi +I) \\
 &&  \hspace{2.2cm}  = ( \hat{q}_{,2} \, \tilde{r}_{,1,-2})_{,-2} - \hat{q}_{,2} \,\tilde{r}_{,1,-2} \, ,  \\
 &&  \hat{q}_t = \hat{q}_{,2} + (\varphi_{,2} - \varphi) \, \hat{q} \, , \qquad
   \tilde{r}_t =  -\tilde{r}_{,-2} - \tilde{r} \, (\varphi_{,2} - \varphi) \, .
\eez   
\end{enumerate}

\subsection{Binary Darboux transformation}
The linear equations (\ref{phi_theta_eq}), (\ref{phi_eta_eq}), (\ref{Omega_Sylvester}) and (\ref{bd_Omega}) 
take the form
\bez
  &&  (\theta_{,1} + \theta \, P)_{,2} 
     = -\theta_{,1}- (\varphi_{0,1}-\varphi_0)_{,2} \, \theta_{,1} \, , \qquad
     \theta_t = \theta_{,2} + (\varphi_{0,2}-\varphi_0) \, \theta \, , \\
  &&  (\eta + Q \, \eta_{,1})_{,-2}  = -\eta  - \eta \, ( \varphi_{0,1}-\varphi_0) \, , \hspace{1.25cm}
     \eta_t = - \eta_{,-2} - \eta \, ( \varphi_{0,2}-\varphi_0) \, , \\
  && \tilde{\Omega}_{,2} - \tilde{\Omega} = \eta \, \theta \, , \quad  
     ( \tilde{\Omega} \, P )_{,2} = Q \, \tilde{\Omega}_{,1} - \eta \, \theta_{,1} + \gamma_1 \, , \quad
     \tilde{\Omega}_t =  \eta_{,-2} \, \theta - \gamma_2 \, . 
\eez
For a given solution $\varphi_0$ of (\ref{2+1_varphi_eq}), we have to find solutions $\theta$ and $\eta$ of 
the first four equations. Then a corresponding solution of the equations for $\tilde{\Omega}$ has to be found. 
According to Theorem~\ref{thm:main_phi},  
\be
    \varphi = \varphi_0 + \theta \, \tilde{\Omega}^{-1} \, \eta_{,-2} \, , \qquad
       \tilde{q} = - \theta \, \tilde{\Omega}^{-1} \, , \qquad 
       \tilde{r} = - \tilde{\Omega}_{,2}^{-1} \, \eta     \label{varphi_sol_3d}
\ee
then solve (\ref{2+1_varphi_eq_scs}) and (\ref{2+1_q,r_eqs}), and thus also the above self-consistent source systems, 
provided the respective condition for $\tilde{\omega}$ is fulfilled. 

\begin{remark}
For constant seed $\varphi_0$ and constant $P,Q$, the linear system for $\theta$ and $\eta$ coincides 
with that in (\ref{sdcm_const_seed_linsys}).
Therefore solutions are given by (\ref{sdcm_cseed_theta,eta_solutions}). Since in the case under consideration  
the above equations for $\tilde{\Omega}$ are equivalent to (\ref{sdcm_Omega}),   
$\tilde{\Omega}$ is given by (\ref{sdcm_cseed_Omega_solution}). Now (\ref{varphi_sol_3d}) provides us 
with an infinite set of solutions of (\ref{2+1_varphi_eq_scs}) and (\ref{2+1_q,r_eqs}).
\end{remark}

\subsection{Self-consistent source extensions of the second family of generalized Volterra lattices} 
Imposing the reduction (\ref{V_shifts}), the systems obtained in Section~\ref{subsec:scs_phi} lead to 
\be
 && (\varphi_{(k)} - \varphi)_t + (\varphi_{(k)} - \varphi +I) ( \varphi_{(l)} - \varphi)_{(k-l)}
    - (\varphi_{(l)} - \varphi)(\varphi_{(k)} - \varphi +I) \nonumber \\
 && \hspace{.8cm}   = ( \hat{q} \, \tilde{r}_{(-l)} )_{(k)} - \hat{q} \, \tilde{r}_{(-l)} \, , \nonumber \\
 && \; \hat{q}_{(k)} = -(\varphi_{(k)} - \varphi +I) \, \hat{q}_{(k-l)} - \hat{q} \, P \, , \nonumber \\ 
 && \tilde{r}_{(-l)} = - \tilde{r} \, (\varphi_{(k)} - \varphi +I) - P \, \tilde{r}_{(k-l)}  \, , 
            \label{pre-gV2_type1}
\ee
respectively 
\be
 && (\varphi_{(k)} - \varphi)_t + (\varphi_{(k)} - \varphi +I) ( \varphi_{(l)} - \varphi)_{(k-l)} 
    - (\varphi_{(l)} - \varphi)(\varphi_{(k)} - \varphi +I)  \nonumber \\
 && \hspace{2.3cm} = ( \hat{q}_{(l)} \, \tilde{r}_{(k-l)})_{(-l)} - \hat{q}_{(l)} \,\tilde{r}_{(k-l)} \, , \nonumber \\
 &&  \hat{q}_t = \hat{q}_{(l)} + (\varphi_{(l)} - \varphi) \, \hat{q} \, , \qquad
   \tilde{r}_t =  -\tilde{r}_{(-l)} - \tilde{r} \, (\varphi_{(l)} - \varphi) \, .
            \label{pre-gV2_type2}
\ee  
Using (\ref{W_from_varphi}), the corresponding self-consistent source extensions of (\ref{genW_eqs}) are 
\be
  && \Big( \sum_{i=0}^{k-1} W_{(i)} \Big)_t + \sum_{i=0}^{k-1} W_{(i)} \sum_{i=k-l}^{k-1} W_{(i)} 
      - \sum_{i=0}^{l-1} W_{(i)} \sum_{i=0}^{k-1} W_{(i)}    \nonumber \\
  && \hspace{.7cm}  = ( \hat{q} \, \tilde{r}_{(-l)} )_{(k)} - \hat{q} \, \tilde{r}_{(-l)} \, , \hspace{8cm} \nonumber \\   
  && \hat{q}_{(k)} = - \sum_{i=0}^{k-1} W_{(i)} \, \hat{q}_{(k-l)} - \hat{q} \, P \, , \quad
     \tilde{r}_{(-l)} = - \tilde{r} \, \sum_{i=0}^{k-1} W_{(i)} - P \, \tilde{r}_{(k-l)}  \, , \label{gV2_type1}
\ee
respectively 
\be
 && \Big( \sum_{i=0}^{k-1} W_{(i)} \Big)_t + \sum_{i=0}^{k-1} W_{(i)} \sum_{i=k-l}^{k-1} W_{(i)} 
        - \sum_{i=0}^{l-1} W_{(i)} \sum_{i=0}^{k-1} W_{(i)} \nonumber \\   
 && \hspace{1.cm}   = ( \hat{q}_{(l)} \, \tilde{r}_{(k-l)})_{(-l)} - \hat{q}_{(l)} \,\tilde{r}_{(k-l)} \, , \nonumber \\
 &&  \hat{q}_t = \hat{q}_{(l)} + \Big( \sum_{i=0}^{l-1} W_{(i)} - \frac{l}{k} I \Big) \, \hat{q} \, , \quad
   \tilde{r}_t =  -\tilde{r}_{(-l)} - \tilde{r} \, \Big( \sum_{i=0}^{l-1} W_{(i)} - \frac{l}{k} I \Big) \, . \hspace{1.2cm}
            \label{gV2_type2}
\ee

For $k=1$ and $l=2$, this yields further variants of a Volterra equation with sources:
\bez
  && W_t + W \, W_{(-1)} - W_{(1)} \, W = ( \hat{q} \, \tilde{r}_{(-2)} )_{(-1)} - \hat{q} \, \tilde{r}_{(-2)} \, , 
            \nonumber \\
  && \hat{q}_{(1)} = - W \, \hat{q}_{(-1)} - \hat{q} \, P   \, , \qquad
     \tilde{r}_{(-2)} = - \tilde{r} \, W - P \, \tilde{r}_{(-1)}) \, ,
\eez
and
\bez
  && W_t + W \, W_{(-1)} - W_{(1)} \, W 
     = ( \hat{q}_{(2)} \, \tilde{r}_{(-1)} )_{(-2)} - \hat{q}_{(2)} \, \tilde{r}_{(-1)} \, , 
            \nonumber \\
  && \hat{q}_t = \hat{q}_{(2)} + (W + W_{(1)} - 2 I) \, \hat{q} \, , \qquad
     \tilde{r}_t = - \tilde{r}_{(-2)} - \tilde{r} \, (W + W_{(1)} - 2 I) \, .
\eez

\begin{remark} 
With constant seed $\varphi_0$, we are led to exactly the same solutions for $\theta, \eta$ and $\tilde{\Omega}$ 
as in Section~\ref{subsec:V1_exact_sol}. Then
\be
    \varphi = \varphi_0 + \theta \, \tilde{\Omega}^{-1} \, \eta_{(-l)} \, , \qquad
    \hat{q} = - \theta \, \tilde{\Omega}^{-1} \, \gamma_i \, , \qquad 
  \tilde{r} = - \tilde{\Omega}_{(l)}^{-1} \, \eta     \label{varphi_sol_2d}
\ee
provides us with solutions of (\ref{pre-gV2_type1}) if $i=2$, and of (\ref{pre-gV2_type2}) if $i=1$. 
Via (\ref{W_from_varphi}) this yields solutions of (\ref{gV2_type1}) and (\ref{gV2_type2}), respectively. 
\end{remark}

\section{Conclusions}
\label{sec:conclusions}
We have shown that the two families of generalized Volterra (or Bogo\-yavlensky) lattice equations 
arise from two semi-discrete integrable equations in three dimensions via reductions. All these equations 
possess a bidifferential calculus formulation that allows the application of general results concerning 
the construction of a binary Darboux transformation and self-consistent source extensions. 

The systems extending generalized Volterra lattice equations to systems with sources, obtained in this work, 
fall into two types. The first type lacks evolution equations for the sources and, moreover, the 
equations constraining sources involve data of the solution-generating method, namely the $n \times n$ matrices 
$P$ or $Q$. This might look strange at first sight, but one has to keep in mind that, if the rank of 
$\omega$ is chosen smaller than $n$, only the corresponding part of those matrices enters the equations 
that govern the effective sources. 
The second type of systems contain evolution equations for all dependent variables. These systems are 
structurally analogous to systems obtained from higher-dimensional integrable systems (like KP) via a 
squared eigenfunction symmetry reduction. 
The method of \cite{CDMH16} generates them without a detour to higher dimensions (which may 
not always be available). 

It would have gone far beyond the scope of this work to analyze all the new systems and the generated families 
of solutions, which include multi-solitons. Only the Volterra and modified Volterra lattice equations 
with sources of the second type have been treated in more detail.

\vskip.1cm
\noindent
\textbf{Acknowledgments.}  
O.C. has been supported by an Alexander von Humboldt fellowship for postdoctoral researchers. 
We have to thank Aristophanes Dimakis for helpful discussions and several contributions to this work. 
We also would like to thank an anonymous referee for suggesting some useful additions to 
the original manuscript.

\end{document}